\documentclass[a4paper]{lmcs}
\pdfoutput=1

\usepackage{lastpage}
\lmcsdoi{21}{3}{28}
\lmcsheading{}{\pageref{LastPage}}{}{}%
{Jul.~03,~2023}{Sep.~18,~2025}{}

\let\divsymb\div
\let\starsymb\star

\usepackage[utf8]{inputenc}
\usepackage{amssymb}
\usepackage{verbatim}
\usepackage{listings}
\usepackage{hyperref}

\newcommand{\citep}[1]{\cite{#1}}
\newcommand{\citet}[1]{\cite{#1}}

\renewcommand{\div}{\divsymb}
\renewcommand{\star}{\starsymb}

\newcommand{\tsum}{{\textstyle\sum}}

\newcommand{\cd}[1]{{\lstinline[basicstyle=\ttfamily]{#1}}}
\lstnewenvironment{code}{}{}
\lstnewenvironment{smallcode}{\lstset{basicstyle=\small\ttfamily\color{darkgreen},identifierstyle=\color{darkgreen}}}{}

\newcommand{\N}{\mathbb{N}}
\newcommand{\Z}{\mathbb{Z}}
\newcommand{\F}{\mathbb{F}}
\newcommand{\Q}{\mathbb{Q}}
\newcommand{\R}{\mathbb{R}}
\newcommand{\K}{\mathbb{K}}
\renewcommand{\S}{\mathbb{S}}
\newcommand{\V}{\mathbb{V}}

\newcommand{\A}{\mathbb{A}}
\newcommand{\X}{\mathbb{X}}

\providecommand{\IF}{\mathbb{I\!F}}
\renewcommand{\IF}{\mathbb{I\!F}}
\renewcommand{\IF}{\mathbb{I}_\mathbb{\F}}

\newcommand{\ul}[1]{\underline{#1}}
\newcommand{\ol}[1]{\overline{#1}}

\newcommand{\hatf}{\hat{f}}
\newcommand{\hatx}{\hat{x}}

\newcommand{\hatD}{\widehat{D}}
\newcommand{\hatX}{\widehat{X}}
\newcommand{\hatY}{\widehat{Y}}

\newcommand{\op}{\mathrm{op}}
\newcommand{\hatop}{\widehat{\op}}

\newcommand{\tp}[1]{\mathcal{#1}}

\newcommand{\tru}{\top}
\newcommand{\fls}{\bot}
\newcommand{\indet}{\uparrow}
\newcommand{\unkwn}{?}

\newcommand{\sumup}{{\sum\makebox[0pt][r]{\raisebox{0.2ex}{\scriptsize$\mathrm{u}\!\;$}}\!\!\,}}
\newcommand{\smfrac}[2]{\mbox{\small$\dfrac{#1}{#2}$}}

\def\@p#1{\mathrel{\ooalign{\hfil$\mapstochar\mkern5mu$\hfil\cr$#1$}}}

\newcommand{\fto}{\rightarrow}
\newcommand{\pfto}{\@p\rightarrow}
\newcommand{\psfto}{\@p\twoheadrightarrow}

\newcommand{\ivl}[1]{{\lfloor #1 \rceil}}
\newcommand{\luivl}[1]{{[ #1 ]}}
\newcommand{\lx}{\underline{x}}
\newcommand{\ux}{\overline{x}}
\newcommand{\ly}{\underline{y}}
\newcommand{\uy}{\overline{y}}

\newcommand{\ax}{\check{x}}
\newcommand{\ay}{\check{y}}
\newcommand{\ex}{e_x}
\newcommand{\ey}{e_y}

\newcommand{\Ariadne}{\textsc{Ariadne}}
\newcommand{\Brausse}{Brau{\ss}e} 
\newcommand{\Konecny}{Kone\u{c}n\'{y}} 
\newcommand{\Mueller}{M\"uller} 

\DeclareMathOperator{\abs}{\mathrm{abs}}
\DeclareMathOperator{\atan}{\mathrm{atan}}
\DeclareMathOperator{\sqrtf}{\mathrm{sqrt}}

\DeclareMathOperator{\convex}{\mathrm{conv}}
\DeclareMathOperator{\dom}{\mathrm{dom}}
\DeclareMathOperator{\range}{\mathrm{range}}

\DeclareMathOperator{\ulp}{\mathrm{ulp}}

\DeclareMathOperator{\inewton}{\mathrm{N}}

\newcommand{\rndd}{\mathrm{d}}
\newcommand{\rndn}{\mathrm{n}}
\newcommand{\rndu}{\mathrm{u}}

\newcommand{\p}{\partial}
\newcommand{\px}{\partial{x}}
\newcommand{\pu}{\partial{u}}

\newcommand{\jacob}{\mathrm{D}}

\newcommand{\id}{\mathrm{id}}

\title{Rigorous Function Calculi in Ariadne}

\author{Pieter Collins\lmcsorcid{0000-0002-8896-9603}}[a]
\author{Luca Geretti\lmcsorcid{0000-0001-6889-0706}}[b]
\author{Sanja Zivanovic Gonzalez}
\author{Davide Bresolin\lmcsorcid{0000-0003-2253-9878}}[c]
\author{Tiziano Villa\lmcsorcid{0000-0002-9671-8804}}[b]

\address{Department of Advanced Computing Sciences, Postbus 616, Maastricht University, 6200MD Maastricht, The Netherlands}
\email{pieter.collins@maastrichtuniversity.nl}

\address{Dipartimento di Informatica, Università degli Studi di Verona, Verona, Italy}
\email{luca.geretti@univr.it, tiziano.villa@univr.it}

\address{Dipartimento di Matematica, Università degli Studi di Padova, Padova, Italy}
\email{davide.bresolin@unipd.it}

\begin{document}

\begin{abstract}
Almost all problems in applied mathematics, including the analysis of dynamical systems, deal with spaces of real-valued functions on Euclidean domains in their formulation and solution.
In this paper, we describe the tool \Ariadne, which provides a rigorous calculus for working with Euclidean functions.
We first introduce the \Ariadne\ framework, which is based on a clean separation of objects as providing exact, effective, validated and approximate information.
We then discuss the function calculus as implemented in \Ariadne, including polynomial function models which are the fundamental class for concrete computations.
We then consider solution of some core problems of functional analysis, namely solution of algebraic equations and differential equations, and briefly discuss their use for the analysis of hybrid systems.
We will give examples of C++ and Python code for performing the various calculations.
Finally, we will discuss progress on extensions, including improvements to the existing function calculus and extensions to more complicated classes of function.
\end{abstract}


\maketitle


\section{Introduction}

Many problems in applied mathematics are formulated in terms of functions.
Examples include trajectories and flow tubes of ordinary differential equations, crossing times for hybrid systems, feedback for control systems, and state spaces and solutions of partial differential equations.
To be able to solve such problems reliably using computational software, we need to be able to work with functions in a \emph{natural}, \emph{rigorous}, and \emph{efficient} way!
Rigour is especially important in mathematical proofs, verification of safety-critical systems, and long chains of reasoning.

The \Ariadne\ software package~\citep{ariadne} provides analysis and verification tools for dynamic systems with continuous state spaces.
This includes nonlinear hybrid systems in which continuous evolution is interspersed with discrete events triggered by conditions on the continuous state.
This is a very general and complex problem, and requires functionality for many different operations, including solution of algebraic and differential equations and of constrained feasibility problems, and requires general-purpose software.
The computational kernel of \Ariadne\ was subsequently developed, based on rigorous numerical methods for working with real numbers, functions and sets in Euclidean space.
Previous publications~\citep{BenvenutiBresolinCasagrandeCollinsFerrariMazziSangiovanniVincentelliVilla2008,CollinsBresolinGerettiVilla2012ADHS} have focused on the use of \Ariadne\ for reachability analysis of systems.
The purpose of this paper is to describe the low-level functionality of \Ariadne\ that has been used to build up the high-level algorithms, with a focus on the support for working with Euclidean functions.
We shall illustrate the functionality with snippets of real C++ code; this code works under \Ariadne~Release~2.5~\citep{ariadne-2.5}.

The computational kernel of \Ariadne\ provides complete support for the operations of rigorous numerics described in Section~\ref{subsec:rigorousnumerics}.
This includes interval arithmetic, solution of linear equations, automatic differentiation, affine and polynomial function models, solution of algebraic and differential equations.
In order to make the package as self-contained as possible, and because existing software libraries did not meet requirements on functionality, it was decided to implement all necessary operations within the package itself (with the exception of the some basic numerical data types, but including double-precision interval arithmetic).
The implementation of these operations in \Ariadne\ is described in \linebreak[0]{Sections~\ref{sec:logicalnumericalalgebraiccalculus}\nobreakdash--\ref{sec:solvingequations}}.
\Ariadne\ also contains support for nonlinear optimisation and feasibility (constraint satisfaction) problems, though these will not be described in this paper.

Perhaps the main innovation of \Ariadne\ is it being designed on the theoretical framework of \emph{computable analysis}, which studies the possibilities and limitations of digital computation for continuous mathematics.
Since continuous mathematics involves spaces of continuum cardinality (notably the Euclidean spaces $\R^n$ and continuous functions $\R^n\to\R^m$), arbitrary objects cannot be described using finite data, but instead require an infinite \emph{stream} of data, like in a decimal expansion.
The computational model is that of Turing machines (or equivalent, such as random access machines) which are the standard accepted model of digital computation, but for which computations run for infinite time, writing infinitely many symbols to the output stream, and potentially using an unbounded amount of memory.
To relate to physically feasible finite-time computation, it is possible to obtain useful information about an object from a finite piece of a data stream defining it.
However, by working directly on mathematical objects, it is much more straightforward to show that an operator is \emph{computable} and develop algorithms to implement it than working directly with floating-point operations.
The representation of spaces of mathematical objects as data streams is associated with a topology on the space.
A fundamental theorem is that any \emph{discontinuous} operator is \emph{uncomputable}, which provides limitations on computation which are also natural from a mathematical viewpoint.
The basic ideas of the theory are described in Section~\ref{subsec:computableanalysis}.

Basing the design of \Ariadne\ on computable analysis has a number of consequences.
We provide data types for mathematical objects such as real numbers and continuous functions, and support the natural computable constructors and operations for these data types, notably comparisons, arithmetic, evaluation and composition.
The data types are abstract interfaces which allow concrete objects to implement core defining operations, such as a real number providing a way to compute dyadic rational bounds to any given accuracy.
In this way, high-level functionality can be easily built-up from low-level operations while preserving the ordinary mathematical semantics.
We call such data types \emph{effective}, distinguishing them from \emph{exact} algebraic objects like rational numbers, but emphasising that they can be effectively computed by Turing machines.
The result of an actual finite computation is a set containing the result, which we call \emph{validated} to emphasise that they are guaranteed to contain the true mathematical result.
Validated objects correspond to finite prefixes of a data stream, which effective objects to the entire stream.
Validated objects support the same algebraic operations as their effective counterparts, and mixed effective-validated binary operations are provided where possible, which ensures that objects provide the expected operations, facilitating programming.
Various \emph{concrete} implementations of validated data types are provided, allowing fine control in the development and implementation of efficient numerical algorithms, while corresponding to the same abstract interface to allow different data types to be used interoperatively.
This conceptual framework is described more fully in Section~\ref{sec:conceptualfoundations}.

Throughout the development process of \Ariadne, we have paid particular attention to software engineering aspects.
The computational kernel of \Ariadne\ is written in standard C++ for efficiency and portability, with a Python interface for easy scripting.
As previously mentioned, basic data types are specified in terms of abstract interfaces, which clarifies the usage of these types, and allows new concrete implementations to be added.
To facilitate using classes from other packages to be easily integrated into \Ariadne, we provide \emph{wrappers} to convert external data types to the \Ariadne\ interfaces.
High level functionality is organised by specifying interfaces for various \emph{solvers}, with each solver being required to implement a closely-related set of operations, such as those related to solving algebraic or differential equations.

In this article we focus on how Euclidean functions are handled in \Ariadne, as this forms the computational backbone of the tool.
For many problems, including as the solution of parametrised algebraic equations, or of differential inclusions, both arguments and results are functions.
Typically, user arguments are given in terms of symbolic formulae, which can be computed arbitrarily accurately.
However, for the results, we use general classes of approximating functions; currently only polynomials are supported in \Ariadne, but one could also consider Fourier series, rational functions and/or splines.
The results are typically only valid over bounded domains.
Arguments and results of solver computations are often functions, so the solver interfaces are tightly bound to the function calculus.

The usage of the function calculus and solvers is illustrated in Section~\ref{sec:examples}.
We consider two problems based on the Fitzhugh-Nagumo system of ordinary differential equations, which arise as a model of the electrophysiology of a squid axon.
This system is nonlinear, so cannot be solved analytically, requiring the use of numerical methods.
We consider two problems, that of computing \emph{fixed-points} of the dynamics, and that of providing a computation of the \emph{flow} of the system over a set of initial states.
These problems illustrate the usage of the solvers for algebraic and differential equations respectively.

The \Ariadne\ tool is still under active development, both in terms of efficiency and usability improvements to existing functionality, and providing new data types and algorithms to handle more complicated classes of functions. We give an overview of some of these extensions in Section~\ref{sec:extensions}.

Finally, we give a brief summary of the paper and more directions for future work in Section~\ref{sec:conclusions}.

\subsection{Other Tools and Approaches}
\label{subsec:tools}

Recent extensions of \Ariadne's function calculus have taken place within the E.U. Horizon Europe project ``Computing with Infinite Data''.
In this project, a number of packages with capabilities for function calculus are being developed:
\begin{itemize}\setlength{\itemsep}{0pt}
\item \cite{ariadne} (Collins, Geretti, Villa et al., 2002--) for verification of hybrid systems~(C++ \& Python). {\small\\\quad\qquad\texttt{https://www.ariadne-cps.org/}}
\item \cite{irram} (\Mueller, \Brausse, 2000--) for real number arithmetic (C++). {\small\\\quad\qquad \texttt{http://irram.uni-trier.de/}}
\item \cite{aern} (\Konecny\ et al.\nocite{DuraczFarjudianKonecnyTaha2014}, 2005--) for effective real computation (Haskell). {\small\\\quad\qquad\texttt{http://michalkonecny.github.io/aern/\_site/}}
\item ERC~(Ziegler, Park et al.\nocite{ParkBrausseCollinsKimKonecnyLeeMuellerNeumannPreiningZiegler202024LMCS}, 2018--), a language for exact real computation.
\end{itemize}
The iRRAM package started off as a tool for efficient and straightforward, high-precision computation with real numbers~\citep{Mueller2001}, but more recent developments have extended the functionality to include a function calculus~\citep{BrausseKorovinaMueller2016} and solution of differential equations.
The AERN package is similar in scope to \Ariadne, including a function calculus~\cite{DuraczFarjudianKonecnyTaha2014}.
The ERC language~\cite{ParkBrausseCollinsKimKonecnyLeeMuellerNeumannPreiningZiegler202024LMCS} has a formal semantics for exact real computation; a motivation for the development of ERC as a computer analysis (as opposed to algebra) system was given in~\cite{BrausseCollinsZiegler2022CASC}.

Other tools for rigorous numerics with a function calculus include:
\begin{itemize}\setlength{\itemsep}{0pt}
\item COSY Infinity~\citep{MakinoBerz2006cosy}, the seminal package in which Taylor models were developed~\cite{MakinoBerz2003}.
\item CAPD Library~\citep{KapelaMrozekWilczakZgliczynski2021capd}, a comprehensive library for analysis of dynamical systems.
\item Flow*~\citep{ChenAbrahamSankaranarayanan2013flowstar}, a tool for hybrid system evolution which also has Taylor models with interval coefficients.
\item JuliaReach~\citep{BogomolovForetsFrehsePotomkinSchilling2019juliareach}, a tool for set-based reachability analysis.
\item DynIBEX~\citep{AlexandreditSandrettoChapoutot2016RC}, has a custom class for trajectories.
\item CORA~\citep{AlthoffGrebenyukKochdumper2018ARCH} has Taylor models,  but operations use floating-point approximations.
\end{itemize}
In particular, COSY Infinity was the first package to implement Taylor models for working with functions, and the the Taylor function calculus of \Ariadne\ is heavily based on the ideas of COSY Infinity.

Other tools have rigorous numerical functionality for differential equations, but lack a full function calculus:
\begin{itemize}\setlength{\itemsep}{0pt}
\item AWA~\citep{Lohner1994awa}
\item VNode~\citep{Nedialkov2006vnode}
\end{itemize}
Finally, we mention some important foundational computational approaches:
\begin{itemize}\setlength{\itemsep}{0pt}
\item Ellipsoidal calculus~\citep{KurzhanskiVaraiya2002I}
\item Set-valued viability~\citep{CardaliaguetQuincampoixSaintPierre1999AISDG}
\item Taylor-ellipsoid models~\citep{HouskaVillanuevaChachuat2013CDC}
\end{itemize}
Since the main aim of this paper is to introduce the framework and capabilities of \Ariadne, we do not give a full comparison with these other tools here.
For a comparison of the performance of \Ariadne\ against other tools on benchmark problems, see~\citep{ARCHCOMPContinuousHybridNonlinear2020,ARCHCOMPContinuousHybridNonlinear2021,ARCHCOMPContinuousHybridNonlinear2022}.

\section{Theory and Methods}
\label{sec:theorymethods}

In this section we give a brief outline of the rigorous numerical methods and computable analysis theory which we will need later.
We first introduce basic techniques of rigorous numerics from the literature which are implemented in \Ariadne.
Details of the algorithms and their implementation will be discussed in Sections~\ref{sec:logicalnumericalalgebraiccalculus}--\ref{sec:solvingequations}.
We then give the basic elements of the theory of computable analysis which we need here, and indicate how it relates to rigorous numerics.
The organisational principles of \Ariadne, which are based on computable analysis, will be described in~Section~\ref{sec:conceptualfoundations}.

\subsection{Rigorous Numerics}
\label{subsec:rigorousnumerics}

The set of real numbers $\R$ and the set of continuous functions $\R^n\rightarrow\R^m$ have continuum cardinality, so there is no possible way of describing all elements exactly using a finite amount of data.
The traditional approaches to handling uncountable sets computationally are either to restrict to a countable subset and use symbolic manipulation, or to work with approximations in a finite subset.
For the purpose of computer-assisted mathematical proofs, or verification of system models, neither approach is feasible; in the former, the computations take far too long, and in the latter, we lose rigour.
Instead, the main idea of rigorous numerics is to work with sets which are guaranteed to contain the correct mathematical value, but which can be expressed with a finite amount of data.

The result of a rigorous numerical computation of an element $x$ from an uncountable set $X$ therefore is a set $\hat{x}$ containing $x$.
The set $\hat{x}$ is taken from a countable set $\hatX$ of subsets of $X$, and has a concrete description given by a finite amount of data.
This is stronger information than that given by traditional approximative numerics, which returns a value $\tilde{x}\in X$ with $\tilde{x}\approx x$, but no guarantees on the approximation quality.

A \emph{validated implementation} of an operation $\op:X_1\times \cdots \times X_\rndn \fto Y$ is a function $\hatop:\hatX_1\times \cdots \times \hatX_\rndn \fto \hatY$ such that the following \emph{inclusion property} holds:
\begin{equation*} \label{eq:validatedinclusion}
\forall i=1,\ldots,n,\ \hat{x}_i \ni x_i \;\implies\; \hatop(\hat{x}_1,\ldots,\hat{x}_\rndn) \ni \op(x_1,\ldots,x_\rndn) .
\end{equation*}

A value $x \in X$ is described by an infinite sequence of \emph{arbitrarily-accurate} over-approximations $\hat{x}_k$ such that for any open $U\ni x$, there exists $n$ such that $\hat{x}_k \subset U$ whenever $k\geq n$.
If $X$ is a Hausdorff space, then $\bigcap_{k=0}^{\infty}\hat{x}_k=\{x\}$ and we write $\lim_{k\to\infty}\hat{x}_k=\{x\}$.
Note that we do not require the sequence to be monotone.
Further a validated implementation $\hatop$ is \emph{arbitrarily accurate} if
\begin{equation*}
 \forall i=1,\ldots,n,\ \lim_{k\to\infty}\hatx_{i,k}=\{x_i\} \;\implies\; \lim_{k\to\infty}\hatop(\hatx_{1,k},\ldots,\hatx_{n,k})=\{\op(x_1,\ldots,x_\rndn)\} .
\end{equation*}
The inclusion property ensures that results are \emph{correct}, and the accuracy property ensures that results are \emph{complete}.

We call the elements of $\widehat{X}$ \emph{basic} sets, since they usually form a basis of open or compact sets of a topological space $X$.
An advantage of using compact sets is that compactness is preserved under continuous forward images.
A secondary advantage is that compact sets include many important singleton constants, which can thus be handled exactly.
In a metric space $(X,d)$ with countable dense subset $\check{X}$, a natural choice for $\widehat{X}$ is the set of \emph{balls} $B(\check{x},e) = \{x\in X \mid d(\check{x},x)\leq e\}$ with $\check{x}\in\check{X}$ and $e\in\Q^+$.
In a partially-ordered space $(X,\leq)$, we may take $\widehat{X}$ to be the set of (generalised) intervals $[\ul{x}\!:\!\ol{x}]=\{x \in X\mid \ul{x}\leq x \leq \ol{x}\}$ with $\ul{x},\ol{x}\in\check{X}$.
We say $\ul{x},\ol{x}$ yield lower- and upper-\emph{bounds} for a value in $X$.

The most well-known example of a rigorous numerical calculus is the \emph{interval arithmetic} of~\citep{Moore1966}; see also~\citep{Neumaier1991,JaulinKiefferDidritWalter2001,MooreKearfottCloud2009}.
A real number $x\in\R$ is represented by an interval $[\ul{x}\!:\!\ol{x}] \!\ni\! x$.
Typically, the endpoints $\ul{x},\ol{x}$ are taken to lie in a finite set $\F$ of \emph{floating-point} numbers of a given precision; we denote by $\IF$ the set of all such intervals.
Alternatively, we take intervals as balls defined by a centre $\check{x}$ and radius $e$, so $x\in \check{x}\pm e$.
An \emph{interval version} $\ivl{\!\star\!}$ of a (binary) operator $\star$ must satisfy the inclusion property
\begin{equation*} x \in [\lx\!:\!\ux] \ \wedge\ y \in [\ly\!:\!\uy] \ \implies \ x \star y \in [\lx\!:\!\ux] \,{\ivl{\!\star\!}}\, [\ly\!:\!\uy] , \end{equation*}
and an \emph{interval extension} of a function $f:\R\fto\R$ is a function $\ivl{f}:\IF\fto\IF$ on intervals satisfying the inclusion property
\begin{equation*}   x\in[\ul{x}\!:\!\ol{x}] \; \implies\; f(x) \in \ivl{f}([\ul{x}\!:\!\ol{x}]) . \end{equation*}
The \emph{natural interval extension} of an arithmetical expression is obtained by using interval arithmetic operations instead of exact arithmetic operations at each stage.
Different expressions for the same function give rise to different natural interval extensions.
These may yield poor approximations due to a phenomenon known as the \emph{dependency effect}:
%
For example, if $f(x)=x(1-x)$, then $\ivl{f}(\luivl{0\!:\!1})=\luivl{0\!:\!1}\times(1-\luivl{0\!:\!1}) = \luivl{0\!:\!1}\times\luivl{0\!:\!1} = \luivl{0\!:\!1}$.
However, $f(x)=x(1-x)=\frac{1}{4}-(\frac{1}{2}-x)^2$, and the natural interval extension of this expression yields $\ivl{f}(\ivl{0\!:\!1})=\frac{1}{4}-(\frac{1}{2}-\luivl{0\!:\!1})^2=\frac{1}{2}-\luivl{-\frac{1}{2}\!:\!+\frac{1}{2}}^2=\frac{1}{4}-\luivl{0\!:\!\frac{1}{4}}=\luivl{0\!:\!\frac{1}{4}}$, a better approximation.

The dependency effect is particularly strong in the use of Gaussian elimination on intervals to solve linear algebraic equations, which yields very poor results for even moderate-size problems.
A viable approach   ~\citep{Rump2010} is to precondition using an approximate inverse $P\approx A^{-1}$ and to use an iterative method such as Gauss-Seidel to solve
\[ PA\;x = Pb . \]

A powerful rigorous calculus for continuous functions $\R^n\fto\R$ is based around the \emph{Taylor models} of~\citet{MakinoBerz2003}.
A Taylor model for a function $f:[-1\!:\!+1]^n\fto\R$ is a pair $\hatf=\langle p,I \rangle$ where $p$ is a polynomial in $n$ variables
\[ p(x)=\sum_{\alpha} c_\alpha x_1^{\alpha_1}\cdots x_\rndn^{\alpha_\rndn} \]
with coefficients $c_\alpha$ in $\F$ and $I$ an interval with endpoints in $\F$ satisfying
\[ \forall z\in[-1\!:\!+1]^n, \ f(z)-p(z) \in I  \]
where $f(z)$ and $p(z)$ are the values obtained using standard real arithmetic.
The interval $I$ is used to capture the round-off errors introduced by floating-point arithmetic, and truncation errors in numerical algorithms.
Alternatively (as in \Ariadne) may replace the interval remainder term with a uniform bound $e$ on the error of the polynomial part.
Taylor models for functions on a general box domain $D=\prod_{i=1}^{n}[a_i\!:\!b_i]$ can be constructed by pre-composing $f$ with $s^{-1}$, where $s$ is a scaling function $[-1\!:\!+1]^n\fto D$.

The usual operations on functions, including arithmetic, evaluation, composition, and antidifferentiation, can be extended to Taylor models, and satisfy the inclusion property.
In principle, differentiation is not possible, since the error between the represented function $f$ and $p$ may have arbitrarily rapid fluctuations, but in practise it is often sufficient to compute the derivative of $p$.
The efficiency of Taylor models relies on \emph{sweeping} terms of $p$ with small coefficients into the interval $I$.

Taylor models for smooth functions can be computed using Taylor's theorem with remainder term.
Automatic differentiation~\citep{Griewank2000} is a technique for computing the derivatives without resorting to symbolic differentiation or divided differences.
The basic idea is to evaluate a formula over a data type containing a value $u$ and its (partial) derivatives $\p{u}/\px_j$ with respect to one or more independent variables.
We then use the basic rules of differentiation to propagate the derivatives through the formula.
From the basic sum, product and chain rules
{\small\[  \frac{\p(u+v)}{\px_j} = \frac{\p u}{\px_j} + \frac{\p v}{\px_j}; \quad \frac{\p(u \times v)}{\px_j} = \frac{\p u}{\px_j} \,  v + u \, \frac{\p v}{\px_j} ; \quad \frac{\p{f(u)}}{\px_j} = f'(u) \, \frac{\pu}{\px_j} . \]}%
we obtain formulae on the automatic differentiation data types.

More complicated operations, notably inverse problems such as the solution of algebraic or differential equations, can be built on top of basic operations on functions.

To solve the algebraic equation $f(y)=0$, a standard technique is to use the \emph{interval Newton} operator~\citep{Moore1966}:
\begin{equation*} \inewton(f,\hat{y},\tilde{y}) =  \tilde{y} - [	\jacob f(\hat{y})]^{-1} f(\tilde{y})  , \end{equation*}
where $\hat{y}$ is a box and $\tilde{y}\in\hat{y}$.
Any solution to $f(y)=0$ in $\hat{y}$ also lies in $\inewton(f,\hat{y},\tilde{y})$, and further, if $\inewton(f,\hat{y},\tilde{y})\subset\hat{y}$, then the equation $f(y)=0$ has a unique solution in $\hat{y}$.
The solution is found by iteratively applying the operator until inclusion is satisfied and the domain is sufficiently small.

There are many methods in the literature for solving differential equations, including~\citep{Lohner1987,BerzMakino1998,NedialkovJacksonCorliss1999,Zgliczynski2002}; see also \citep{Tucker2011}.
However, most use some variation on the following procedure to compute the flow $\phi$ of $\dot{x}=f(x)$ starting in an initial set $X_0$ at time $0$ for a time step $h$.
The flow is computable as a fixed-point of the Picard operator
\[ \phi(x,t) = x + \textstyle{\int_{0}^{t}} f(\phi(x,\tau) \mathrm{d}\tau .  \]
First a \emph{bound} $B$ is computed such that $\phi(X_0,[0,h])\subset B$.
A sufficient condition for $B$ to be such a bound is that $X_0+[0,h]f(B) \subset B$.
Either the Picard operator is used directly, or to find the Taylor coefficients of $\phi$ using automatic differentiation, both at the point $(x_0,0)$ and over the set $(B,[0,h])$.
Finally, the results are combined to give a set $X_1$ containing $\phi(X_0,h)$.
As an additional step, we may apply a \emph{reconditioning} operation to control the complexity of the representation.

Other problems which have been well-studied in the literature include nonlinear programming~\citep{HansenWalster2004} and constraint propagation~\citep{Kearfott1996,JaulinKiefferDidritWalter2001}.
The latter is used in \Ariadne\ to test emptiness of the intersection of sets.

\subsection{Computable Analysis}
\label{subsec:computableanalysis}

Computable analysis~\citep{Weihrauch2000} provides a theoretical foundation for rigorous numerics.
An operation is \emph{computable} if it the result can be rigorously computed to arbitrary accuracy.

Mathematical objects from sets of continuum cardinality, such as numbers, functions and sets, are described by infinite \emph{streams} of data $\{0,1\}^\omega$, in such a way that useful information is provided by a finite part of the stream.
Formally, a \emph{representation} of as set $X$ is a partial surjective functions $\delta:\{0,1\}^\omega\psfto X$, and an element $p\in\{0,1\}^\omega$ such that $\delta(p)=x$ is a \emph{$\delta$-name} of $x$.
Not all representations are useful; those that are are called \emph{admissible}.

Computations on represented sets are described by Turing machines (finite programs) acting on data streams encoding names.
Since computations output an infinite stream of data, they should not terminate, and may use an infinite amount of internal memory (though memory use after a finite time must be finite).
An operator $\op:X_1\times\cdots\times X_\rndn \to Y$ is \emph{computable} with respect to representations on the inputs and outputs if it can be implemented by a Turing machine.
An element $x:X$ is computable if a name can be computed by a Turing machine with no inputs.

A set may have many admissible respresentations.
Two representations are \emph{equivalent} if it is possible to compute a name for an object in either representation from any name in the other.
Sets with equivalence classes of admissible representations we call a \emph{(computable) type}
To emphasise that these types correspond to the standard mathematical spaces, we will sometimes use the terminology \emph{mathematical type} instead.

Representations of mathematical types are strongly related to topologies on the represented sets.
Well-behaved representations of a topological space are quotient maps, and are called \emph{admissible quotient representations} in~\citep{BattenfeldSchroederSimpson2007ENTCS}.
The topological spaces arising in analysis and geometry all have a unique ``natural'' representation up to equivalence, so have a canonical computable type.
(An ``unnatural'' representation of the real numbers can be obtained by shifting by an uncomputable constant, but arithmetic is then uncomputable with respect to this representation.)
A fundamental result is than only \emph{continuous} operators can be \emph{computable}.
(Further, we do not know of any ``natural'' continuous operators which are uncomputable.)

One of the main properties of computable analysis is that comparison of two real numbers is undecidable; if $x,y\in\R$ and $x\neq y$ then we can indeed prove either $x<y$ or $x>y$ in finite time, but if $x=y$, then we will never know this.
Even for the case of symbolic numbers defined using arithmetic, exponential and trigonometric functions, it is unknown whether equality is decidable.
To ensure all decision computations are finite, we allow such comparisons to return a value ``indeterminate'', which contains no information.
The resulting logical type we call the \emph{Kleenean}s, denoted $\K$.

Functions are \emph{defined} by evaluation.
Thus a name of a real function $f:\R\fto\R$ is a data stream which, given a name of a real number $x$, yields a name of the real number $f(x)$.
It turns out that this representation is equivalent to providing an interval extension allowing arbitrarily accurate evaluation, as defined in Section~\ref{subsec:rigorousnumerics}.
This gives the fundamental link between computable analysis and rigorous numerics: a computable function has an arbitrarily accurate validated implementation, while providing such an implementation is a proof of computability.

The most important properties of computable analysis are:
\begin{description}
\item[Canonical types] There are unique canonical types for $\mathbb{B}$ (Booleans), $\mathbb{S}$ (Sierpinskians), $\mathbb{K}$ (Kleeneans), $\N$, $\Z$, $\Q$ and $\R$.
\par
The \emph{Kleenean} type $\mathbb{K}$ has values $\{\top,\bot,\indet\}$ where $\indet$ (``indeterminate'') indicates that a proposition is undecidable.
It has Kleene semantics, with implication $\indet \;\rightarrow\; \indet \;=\; \indet$.
Note that $\indet$ is not considered an ``error'', since any nontrivial predicate on continuous data has undecidable instances.
$\mathbb{S}$~is the subtype with values $\{\top,\indet\}$.
\par
The real type $\mathbb{R}$ is the unique type of the set of real numbers supporting arithmetic, comparisons, and limits of strongly-convergent Cauchy sequences. Comparisons yield values in $\K$, with $x < y = x \leq y =\; \indet$ if $x=y$.
\par
Any (effectively) separable complete metric space has an admissible quotient representation given by Cauchy sequences $(x_\rndn)$ which are fast-converging: $d(x_m,x_\rndn)\leq 2^{-\min(m,n)}$.
\par
\item[Quotients of countably-based spaces] A topological space $X$ has an admissible quotient representation if, and only if, it is the quotient space of a countably-based space.
\par
\item[Products] Given types $\tp{X}_1$ and $\tp{X}_2$, the  \emph{product} type $\tp{X}_1\times\tp{X}_2$ exists, with computable projection functions $\pi_i:\tp{X}_1\times\tp{X}_2\fto\tp{X}_i$.
Infinite products $\prod_{i=0}^{\infty}\tp{X}_i$ also exist.
\item[Functions] Given types $\tp{X}$ and $\tp{Y}$, the  \emph{continuous function} type $\tp{C}(\tp{X};\tp{Y})$ exists, with computable \emph{evaluation} $\varepsilon:(\tp{C}(\tp{X};\tp{Y})\times \tp{X} \to \tp{Y}$. There is a  computable natural bijection between $(\tp{X}\times\tp{Y})\fto\tp{Z}$ and $\tp{X}\fto \tp{C}(\tp{Y};\tp{Z})$.
\par
A function $f:\tp{X}\to\tp{Y}$ is computable if, and only if, it has a computable name as an element of $\tp{C}(\tp{X};\tp{Y})$. For this reason, we shall henceforth use $\tp{X}\to\tp{Y}$ synonymously with $\tp{C}(\tp{X};\tp{Y})$.
\item[Subtypes] Any subset of a type is itself a type.
\item[Derived types] We can easily construct types from other types.
\par
For example, open subsets of a space $X$ are described by the type $\tp{X}\fto\mathbb{S}$, since a set $U$ is open if, and only if, its indicator function $\chi_U:X\fto\mathbb{S}$ is continuous.
Similarly, compact sets are described by $(\tp{X}\fto\mathbb{S})\fto\mathbb{S}$ using the predicate $C(U) \iff C\subset U$.
\end{description}
Since we can construct product and function types, computable types form a \emph{Cartesian closed category}, which allows constructions of the \emph{lambda-calculus}; see~\citep{LambekScott1988}.

For a more complete introduction to computable analysis, see~\citep{Weihrauch2000,Collins2020MSCS}.

\definecolor{darkorange}{rgb}{0.5,0.25,0.0}
\definecolor{orange}{rgb}{1.0,0.5,0.0}

\newif\ifcoloranglebrackets
\coloranglebracketsfalse
\newcommand\coloroncondition{\ifcoloranglebrackets\color{orange}\fi}

\definecolor{darkred}{rgb}{0.5,0.0,0.0}
\definecolor{darkgreen}{rgb}{0.0,0.5,0.0}
\definecolor{darkblue}{rgb}{0.0,0.0,0.5}
\lstdefinestyle{ariadnestyle}{
    basicstyle=\ttfamily\color{black},
    morekeywords={requires},
    emph={[1]N,I,X,XE,X1,X2,FLT,F,FE,PR,PRE,P,SIG,RES,ARG,ARGS,ARGS...},emphstyle={[1]\color{cyan}},
    numberstyle=\color{blue},
    identifierstyle=\color{magenta},
    stringstyle=\color{red},
}

\lstset{language=C++,style=ariadnestyle}

\section{Conceptual foundations}
\label{sec:conceptualfoundations}

In \Ariadne, we identify three orthogonal concepts in the abstract description of objects, namely the mathematical \emph{type}, the kind of \emph{information} provided (both related to ideas from computable analysis described in Section~\ref{subsec:computableanalysis}), and a distinction between \emph{generic} objects, and \emph{concrete} objects defined by \emph{properties} affecting the available precision.
In this section, we introduce these concepts, and how they are described in \Ariadne\ C++ code.

\subsection{Introduction to C++ Syntax}

In C++, data types are called \emph{classes}.
In \Ariadne, we use \cd{CamelCaps} as class names.
C++ supports \emph{dependent} types as \emph{templates}, so \cd{Float<PR>} denotes a floating-point number class with precision given by the template parameter \cd{PR}, and \cd{Float<DoublePrecision>} denotes an instantiation using double-precision.
In \Ariadne, we use uppercase abbreviations \cd{CAPS} as template parameters.
C++ classes can be given aliases, so \cd{FloatDP} is an alias for \cd{Float<DoublePrecision>}.
Also, such aliases can be local to a class, so \cd{Float<PR>::PrecisionType} is an alias for the actual type used for the precision.
Such aliases can also be provided as templates, such as \cd{PrecisionType<F>}, with \cd{PrecisionType<Float<DoublePrecision>>} being an alias of \cd{Float<DoublePrecision>::PrecisionType}.

Functions in C++ are are specified by their \emph{signature}, which is given in a declaration such as
\cd{add(FloatDP x1, FloatDP x2) -> Bounds<FloatDP>}.
The argument names may be omitted, giving \cd{add(FloatDP, FloatDP) -> Bounds<FloatDP>}.
Infix operators are declared by using the syntax \cd{operator+(FloatDP,FloatDP) -> Bounds<FloatDP>}, but ``\cd{operator}'' is omitted when the operator is called as in \cd{x1 + x2}.
A function associated with a class is a \emph{method}, declared using the syntax \cd{Float<PR>::precision() -> PR}, and called as \cd{x.precision()}.

\subsection{Types}

A \emph{type} in \Ariadne\ corresponds to a mathematical (computable) type of computable analysis.
For example, the type \cd{Real} corresponds to the mathematical real numbers.
The type defines the permitted operations; for example, we can add two real numbers, $\R+\R\fto\R$, compare real numbers $\R\lesssim\R\fto\K$, where $\K$ is the \cd{Kleenean} type representing the result of quasidecidable propositions, and convert from a rational $\Q\hookrightarrow\R$.

Since classes such as \cd{Real} can support arbitrary objects of the mathematical type with the corresponding information we require the use of polymorphism and virtual methods for the operations provided.
The virtual methods required are specified in an \emph{interface} class. For example, we can compute dyadic bounds on a real number to a given accuracy:
\begin{code}
    Real::Interface::compute(Accuracy acc) -> Bounds<Dyadic>;
\end{code}
This is dispatched from the \cd{Real} class directly
\begin{code}
    Real r=sin(1); r.compute(acc);
\end{code}

A related type is the class \cd{UpperReal}, denoted $\R_>$.
While a real number can be bounded arbitrarily accurately from above and below, an upper real only contains enough information to compute upper bounds.
Hence the predicate $\R_> < \Q \fto \S$, where $\S$ is the \emph{Sierpinskian} type, is computable, since we can verify $x<q$ for $x\in\R_>$ and $q\in\Q$, but we cannot verify $x>q$.
Upper reals can be added $\R_>+\R_>\fto\R_>$ but only positive upper reals multiplied ${\R^+_>}\times{\R^+_>}\fto{\R^+_>}$.
Note $\R^+ \times \R_> \fto \R_>$.
As a topological space, the upper reals are non-Hausdorff, and have basis $\{(-\infty\!:\!q) \mid q\in\Q\}$.
An important usage of the positive upper reals is as the radius of open balls in metric spaces.

Many classes represent mutually-incompatible sets of objects.
For example, the class \cd{Vector<Real>} has a \cd{size()} method, and arithmetic cannot be applied to vectors with different sizes.
In order to construct a zero vector, we need to specify the \cd{size}; in particular, vectors do not have a natural default constructor.
The information needed to construct a null element of a class are called \emph{characteristics}, and \cd{CharacteristicsType<Y>} is a tuple class containing this data.
Numerical classes such as \cd{Real} have no characteristics.

\subsection{Information}

In \Ariadne, every object indicates the kind of \emph{information} that object provides about the value of a quantity.

With \emph{exact} information, objects are specified with enough information to identify them exactly; further, equality on such objects is decidable.
For example, $2/3$ is an exact \cd{Rational} number, and $0.625=0.101_2$ is an exact \cd{Dyadic} number.

With \emph{effective} information, objects are specified by a way of finding them to arbitrary precision.
This corresponds to a description by a data stream from computable analysis, or an algorithm computing such a stream (for a computable object).
For example, an algorithm computing guaranteed bounds on the \cd{Real} number $\exp(x)$ for rational $x$ is $\exp(x)=\sum_{n=0}^{N-1} x^n/n! \pm 2 |x|^N/N!$ whenever $N\geq2|x|$.
The \cd{UpperReal} class also provides effective information, but only allows the computation of guaranteed upper-bounds.

An important case of effective information is \emph{symbolic} information.
Objects are specified by symbolic formulae with enough information to identify them exactly, but for which equality is not necessarily decidable.
For example, $2/3$, $\pi$ and $\sin(1)$ are all symbolic reals, and $\exp(1)$ is an symbolic representations of $e$.
For symbolic objects to be effective we need to give an algorithm computing them, and \Ariadne\ provides a default computation in this case.

With \emph{validated} information, objects are specified with sufficient information is given to compute all operations to some fixed accuracy.
For example, a double-precision floating-point interval $[\ul{x}\!:\!\ol{x}]$ is a concrete data type \cd{Bounds<FloatDP>} for the real numbers in the validated paradigm, representing all real numbers $x$ with $\ul{x}\leq x \leq \ol{x}$, and the interval $[2.625\!:\!2.75]$ is a validated representation of $e$.
Similarly, an interval $(\,:\!\ol{x}]$ is a concrete data type \cd{UpperBounds<FloatDP>} for the upper real numbers, with $(\,:\!\ol{x}]$ representing all numbers $x\leq\ol{x}$.

An important case of validated information is a finite prefix of a data stream giving effective information. For example, from $\pi=3.14159\cdots$ we can deduce $\pi\in3.141[59\!:60]$.
There is no \emph{canonical} conversion from effecive to validated information, since we need to specify some accuracy, but we can often perform a binary operation on an effective and a validated object, yielding a validated object.
For example, given $\hat{\pi}=3.141[59\!:60]$ and $\mathrm{e}$, we can approximate $\mathrm{e}$ to $5$ decimal places, yielding $\hat{\pi}+\mathrm{e}=3.141[59\!:60]+2.7182[8\!:9]=5.3134[1\!:\!3]$.

With \emph{approximate} information, a single value is given for a quantity with no guarantees on the exact value of the quantity.
For example, $2.75$ of type \cd{Approximation<FloatDP>} is a floating-point approximation to $e$, with relatively poor accuracy.

It should be noted that a conversion is \emph{safe} if, and only if, the corresponding mathematical types can be converted and the information is weaker.
Hence an exact element of $\R$ can be safely converted to a validated element of $\R_>$, but an exact element of $\R_>$ should not be converted to a validated element of $\R$ i.e. an interval (though the conversion could in principle return an interval $[-\infty,\ol{x}]$).

It is sometimes useful to convert an approximate object to an exact object. In this case the conversion requires an explicit \texttt{cast\_exact}, and it is the responsibility of the user to ensure that the resulting computation is rigorous. An example is using an approximate matrix inverse $\widetilde{A^{-1}}$ as a preconditioner $P$ in solving the linear system $PAx = Pb$.

The kind of \emph{information} about an object was previously called the \emph{(computational) paradigm} used.
In \Ariadne, the type alias \cd{Paradigm<X>} is a tag describing the information of class \cd{X}, either \cd{ExactTag}, \cd{EffectiveTag}, \cd{ValidatedTag} or \cd{ApproximateTag}.
For example, \cd{Paradigm<Real>} is \cd{EffectiveTag}, \cd{Paradigm<Rational>} and \cd{Paradigm<FloatDP>} are \cd{ExactTag}, and \cd{Paradigm<Bounds<FloatDP>>} is \cd{ValidatedTag}.
Since classes of providing different information about the same mathematical type support the same operations, many \Ariadne\ classes are templated on the information tag, for which we use template parameter \cd{P}.
For example, \cd{Number<P>} is a class providing information of kind \cd{P} about a real number.
For such classes, non-templated synonyms are provided, so for example, \cd{EffectiveNumber} is a synonym for \cd{Number<EffectiveTag>}.

\subsection{Properties}

In \Ariadne\ we distinguish between \emph{generic} and \emph{concrete} objects, so that e.g. \cd{FloatMPBounds} is a concrete description of a generic \cd{ValidatedNumber}.
Generic classes \cd{Real} or \cd{ValidatedNumber} require the use of polymorphism and virtual functions for the operations provided.
However, for efficient computation, the need for virtual dispatching of every operation is inefficient.
For this reason, \Ariadne\ supports concrete objects for actual computations.

Every concrete object has an underlying mathematical type (such as \cd{Real}) and information (typically \cd{Exact}, \cd{Validated} or \cd{Approximate}), and is further specified by \emph{properties} giving computational parameters.
Any floating-point number has a \emph{precision} property, with the \cd{FloatMP::precision()} method returning a \cd{MultiplePrecision} (or \cd{MP}) value.
The concrete multiple-precision floating-point numbers $\F$ are \emph{graded} by the precision parameter $p$, so $\F=\bigcup_{p\in P} \F_p$.
Any concrete object of class \cd{X} can be constructed by giving a value of \cd{GenericType<X>} and a tuple of \cd{PropertyType<X>}.
The \cd{CharacteristicsType<X>} for a concrete object is then product of \cd{CharacteristicsType<GenericType<X>>} and \cd{PropertiesType<X>}.

\section{Core Functionality: Logic, Numbers, Algebra}
\label{sec:logicalnumericalalgebraiccalculus}

In this section, we give an overview of the implementation of core logical, numerical and algebraic data types of \Ariadne.
The algorithms are mostly trivial, so we focus on the supported operations and how the implementation relates to the concepts of Section~\ref{sec:conceptualfoundations}.
As well as being important to users in their own right, these are crucial for implementing the function calculus described in Section~\ref{sec:functioncalculus}.

\subsection{Logic}
\label{subsec:logic}

\Ariadne's core logical type is the \cd{Kleenean} type $\mathbb{K}$, which represents the result of a quasidecidable predicate, notably comparison of two real numbers $\R < \R \fto \K$, which has C++ signature
\begin{code}
    operator<(Real, Real) -> Kleenean;
\end{code}
The Kleenean type supports negation $\neg$, conjunction $\wedge$, and disjunction $\vee$, represented by the usual C++ operators \cd{!}, \cd{\&\&} and \cd{||}.

Although a Kleenean represents a value which must be either true $\tru$, false $\fls$, or indeterminate~$\indet$, its representation is not that of the three-point discrete space $\{\tru,\indet,\fls\}$.
If we consider how to compare two real numbers, we may proceed by computing each to $n$ decimal places of accuracy for increasingly large $n$.
If we find $x\in[3.140\!:\!3.143]$ and $y\in[3.142\!:\!3.144]$ we cannot deduce $x<y$ or $y<x$, but need to compute to some higher, and a-priori unknown, accuracy.
Hence, $\K$ is represented by sequences of discrete values $\{\tru,\unkwn,\fls\}$, where `$\unkwn$' denotes an ``unknown'' indeterminate, and no sequence contains contradictory information $\tru$ and $\fls$.
The comparison $3.14159\cdots < 3.14285\cdots$ might therefore yield the sequence $(\unkwn,\,\unkwn,\unkwn,\unkwn,\tru,\tru,\ldots)$ indicating that we need $4$ decimal places to determine the comparison.

In \Ariadne, the \cd{ValidatedKleenean} class has one of the three values \cd{TRUE}, \cd{FALSE} or \cd{INDETERMINATE} (or \cd{indeterminate}).
The \cd{Kleenean} class is defined by the method
\begin{code}
    Kleenean::check(Effort) -> ValidatedKleenean;
\end{code}
where \cd{Effort} is a positive integer which is an abstraction of the amount of work used to check the value.

To be used in a C++ conditional, one needs to convert an \Ariadne\ logical value to a value of the builtin type \cd{bool}, which can be done using
\begin{code}
    definitely(ValidatedKleenean k) -> bool;
    possibly(ValidatedKleenean k) -> bool;
\end{code}
where \cd{definitely} returns \cd{false} if \cd{k} has value \cd{INDETERMINATE}, whereas \cd{possibly} returns \cd{true}.

Comparing \cd{Validated} numbers directly gives \cd{Validated} logical types:
\begin{code}
    ValidatedKleenean operator<(ValidatedReal, ValidatedReal);
    ValidatedKleenean operator<(FloatMPBounds, FloatMPBounds);
\end{code}

Comparing \cd{Approximate} numbers gives a \cd{ApproximateKleenean}, with values \cd{LIKELY} and \cd{UNLIKELY}, which can be converted to a \cd{bool} using the \cd{probably} function.
Note that in the case of the approximate information, the values \cd{TRUE} and \cd{FALSE} are never used, since if $\tilde{x}\approx x$ and $\tilde{y}\approx y$, but no error bounds are provided, we cannot deduce $x>y$ even if $\tilde{x}>\tilde{y}$; in this case the result \cd{LIKELY} is used.

\Ariadne\ also provides a \cd{Boolean} type for the result of decidable predicates.

Any logical object can be coerced into a \cd{bool} for use in a Boolean context (such as an \cd{if} or \cd{while} condition) using the \cd{decide} predicate,
The result implementation-defined on an \cd{INDETERMINATE} value.

\subsection{Numbers}
\label{subsec:numbers}

In \Ariadne, we provide algebraic \cd{Integer}, \cd{Dyadic} and \cd{Rational} classes, based on the GMP library~\citep{gmp}.
Classes \cd{Nat}, \cd{Nat32}, \cd{Int} and \cd{Int32} are wrappers for C++ builtin integral types, and a \cd{Decimal} class is provided for user input.

Real numbers are represented by the \cd{Real} class, which internally uses a symbolic description.
The standard arithmetical operators $+,-,\times,\div$ are supported (in code as \cd{+}, \cd{-}, \cd{*}, \cd{/}), as are the following named elementary functions:
\[ \begin{gathered}
  \mathrm{nul}(x)=0, \quad \mathrm{pos}(x)=+x, \quad  \mathrm{neg}(x)=-x, \quad \mathrm{sqr}(x)=x^2, \quad \mathrm{hlf}(x)=x / 2, \quad \mathrm{rec}(x)=1/x; \\
  \mathrm{add}(x_1,x_2)=x_1\!+\!x_2, \ \  \mathrm{sub}(x_1,x_2)=x_1\!-\!x_2, \ \  \mathrm{mul}(x_1,x_2)=x1\!\times\!x_2, \ \  \mathrm{div}(x_1,x_2)=x_1\!\div\!x_2 ; \\
  \mathrm{pow}(x,n)=x^n, \quad \mathrm{fma}(x_1,x_2,y)=(x_1\times x_2)+y; \\[\jot]
  \sqrtf(x), \ \exp(x), \ \log(x), \ \sin(x), \ \cos(x),\ \tan(x), \ \atan(x) ; \\
  \abs(x)=|x|, \quad \mathrm{max}(x_1,x_2), \quad \mathrm{min}(x_1,x_2);\\
\end{gathered} \]

The generic \cd{Number<P>} template supports the standard information tags, with type aliases e.g. \cd{EffectiveNumber} for \cd{Number<EffectiveTag>}.
Each of these classes supports the same operations as \cd{Real}.
\begin{rem}
The classes \cd{Real} and \cd{EffectiveNumber} are mathematically equivalent, and may be combined in a future release.
\end{rem}

Concrete representations of numbers for use in computations are defined using fundamental floating-point types, of which \cd{FloatDP} (based on C++ \cd{double}) and \cd{FloatMP} (based on the \cd{mpfr\_t} of the MPFR library~\citep{mpfr}) are currently supported.
These classes provide \emph{rounded} operations.
Rounded versions of standard arithmetical operations $\star\in\{+,-,\times,\div\}$ on $\F$ satisfy:
\begin{equation*}\begin{gathered}
    x \star_\rndd y \leq x \star y \leq x \star_\rndu y , \\[\jot]
 \forall z\in\F, \ |x \star_\rndn y - x \star y| \ \leq\  | z - x \star y| ;
\end{gathered}\end{equation*}
where $\star_\rndd$, $\star_\rndu$ and $\star_\rndn$ are, respectively, downwards, upwards, and to nearest rounded versions of $\star$.
Error bounds on operations easily be computed:
\begin{equation*}
 e_\star(x,y) := |(x \star y) - (x \star_\rndn y)| \ \leq \; \bigl (x \star_\rndu y) -_\rndu (x \star_\rndd y) \bigr) / 2 \leq \tfrac{1}{2} \ulp(x\star y) .
\end{equation*}
Here $\ulp(z)$ is the number of \emph{units in the last place} of $z$, defined as
\begin{equation*}
 \ulp(z) := 2^{-(\lfloor \log_2 z \rfloor + d)}
\end{equation*}
where $d$ is the number of binary digits in the mantissa of $z$.
Note that unary $+$ and $-$, as well as the absolute value $|\cdot|$, can be computed exactly for floating-point types.
Modern microprocessors implement rounded operations for the C++ built-in \cd{double} type.
In \Ariadne, rounded operations on floating-point class \cd{F} have the signature
\begin{code}
    op(RoundingMode rnd, F x) -> F
\end{code}
where \cd{rnd} can be one of \cd{down},\cd{up} or \cd{near}.
Hence \cd{mul(down,x1,x2)} computes $x_1 \times_\rndd x_2$.
\begin{rem}
In a future release, we plan to change the signature of rounded arithmetic to
\begin{code}
    op(RoundingMode rnd, F x) -> Rounded<F>}
\end{code}
to indicate that the result is not exact.
\end{rem}

The raw rounded operations are not intended to be directly called in high-level user code, but are used in a safe way in the validated \cd{Bounds<F>} and \cd{Ball<F,FE>} classes, and an approximate \cd{Approximation<F>} class.

The \cd{Bounds<F>} class stores a lower-bound $\lx$ and upper-bound $\ux$ a value $x$, which are accessed using
\begin{code}
    Bounds<F>::lower() -> LowerBound<F>;
    Bounds<F>::upper() -> UpperBound<F>;
\end{code}

Arithmetic on \cd{Bounds<F>} and \cd{Ball<F,FE>} is standard \emph{interval arithmetic} (see~\citep{MooreKearfottCloud2009}), which can easily be implemented in terms of rounded arithmetic.
\begin{equation*} [\lx\!:\!\ux]-[\ly\!:\!\uy] = [ \lx -_\rndd \uy  :  \ux -_\rndu \ly] \end{equation*}
\begin{equation*} (\ax\pm\ex) \times (\ay\pm\ey) = (\ax \times_\rndn \ay) \pm \bigl( \varepsilon_\times(\ax,\ay) \,+_\rndu\, \ex\times_\rndu\ey \,+_\rndu\, |\ax| \times_\rndu \ey \,+_\rndu\, \ex \times_\rndu |\ay| \bigr).  \end{equation*}
Here, $\varepsilon_\times(\ax,\ay)$ is an over-approximation to $|\ax\times_\rndn \ay - \ax\times\ay|$. Note that $|x|$ can be evaluated exactly.

\begin{rem}[Terminology]
In \Ariadne, we use \cd{Bounds} as the name of the class used for interval arithmetic, since the \cd{Interval} template class in \Ariadne\ exclusively refers to a \emph{geometric} interval representing a mathematical set $[a\!:\!b]$.
This differs from the meaning of \cd{Bounds} to refer to the possible values $[\ul{x}\!:\!\ol{x}]$ of a single \emph{numerical} quantity $x$.

A geometric interval with upper endpoint of type \cd{UB} is represented by the class \cd{Interval<UB>}.
Note that while the data $\langle \ul{a},\ol{b} \rangle$ for an over-approximation to an interval $[a\!:\!b]$ is the same as that of an over-approximation $\langle \ul{x},\ol{x} \rangle$ to a single point $x$, the meaning is subtly different.
\end{rem}

To facilitate user calculuations, \Ariadne\ provides mixed binary arithmetical operations.
The information provided by the result type must be appropriate, typically the weaker of the two arguments, and mixed operations involving a generic object and a concrete object yield a concrete object.
Where possible, these operations use C++ type conversion rather than an explicit implementation.
The signature of some illustrative mixed operations are
\begin{code}
    operator+(Ball<F,FE>, F) -> Ball<F,FE>;
    operator-(Real, LowerBound<F>) -> UpperBound<F>;
    operator*(ValidatedNumber, Float<PR>) -> Bounds<Float<PR>>;
\end{code}
The templated type alias \cd{ArithmeticType<X1,X2>} is defined as the result of the expression \cd{x1 * x2 - x1 * x2}, where \cd{x1},\cd{x2} have types \cd{X1},\cd{X2} respectively.

While C++ is a powerful and expressive language, it was not designed with numerical rigour in mind, and contains features which may result in unsafe computations, notably relating to the built-in types. We now briefly describe two of these features, namely automatic conversions and floating-point literals, and how we restore safe usage in \Ariadne.

\begin{rem}[Conversions]
\label{rem:conversions}
The builtin types, including \cd{bool}, \cd{char}, \cd{unsigned int}, \cd{int}, \cd{long int}, \cd{float} and \cd{double}, may be automatically converted to each other.
This is particularly dangerous since floating-point types may be converted to integral types, or integral types may be converted to others with a smaller number of bits or different sign.
For this reason, \Ariadne\ does not support mixed operations involving \Ariadne\ classes and builtin numbers, or conversion from builtin numbers to \Ariadne\ classes, unless these operations are safe.
The main restriction is that builtin floating-point numbers can only be used where \Ariadne\ would accept a real number carrying \cd{Approximate} information.
\par
The allowed arguments of template methods can be restricted using \emph{concepts} from the C++20 standard, as in the following constructor of \cd{Integer}:
\begin{code}
    template<class N> requires Integral<N> Integer(N const& n);
\end{code}
\end{rem}

\begin{rem}[Literals]
\label{rem:floatliterals}

The C++11 standard allows for user-defined integer, floating-point and string literals.
This allows for disambiguation of the meaning of a floating-point number in text.

The \Ariadne\ literal suffix \cd{\_x} denotes an \emph{exact} floating-point literal.
Hence \cd{r=0.25\_x} explicitly states that the literal \cd{0.25} denotes a the number $1/4$, represented as a floating-point number.
An error occurs if the value is not exactly representable as a double-precision floating-point number, such as \cd{0.3\_x}.
(This is tested for by checking whether the input value, which is a C++ \cd{long double}, retains its value when converted to a \cd{double}.)

The suffix \cd{\_pr} denotes a sufficiently \emph{precise} floating-point number.
Hence \cd{r=0.3\_x} means that $r$ is a floating-point number with value given by the \emph{compiler's approximation of the string ``\cd{0.3}'' as a floating-point number}.
The value of this number is
\[ 0.299999999999999988897769753748434595763683319091796875 \text{\ .} \]

A \emph{rational literal} has subscript \cd{\_q}.
A floating-point value is converted to a rational using continued fractions.
If any iterate of the exact continued-fraction expansion $a_{i+1}=1/(a_i-\lfloor a_i\rfloor)$ has value less than $\epsilon$, this is assumed to be a rounding error, and the value is set to $0$.

A \emph{decimal literal} has subscript \cd{\_d} or \cd{\_dec}.
The nearest decimal number with at most $s$ significant digits is computed.
If this has smaller error than machine epsilon $\epsilon$, the decimal approximation is accepted.
\end{rem}

\newcommand{\tRING}{A{\small RING}}
\newcommand{\tALGEBRA}{A{\small LGEBRA}}
\newcommand{\tBANACH}{B{\small ANACH}}
\newcommand{\tGRADED}{G{\small RADED}}
\newcommand{\tDIFFERENTIAL}{D{\small IFFERENTIAL}}

\subsection{Linear algebra}
\label{subsec:linearalgebra}

The generic \cd{Vector<X>} class represents vectors over a (numeric) field, or a module over an algebra, with values of type \cd{X}.
The value type is accessible as \cd{Vector<X>::ScalarType} and the type of the underlying field (if \cd{X} is an algebra; see Section~\ref{subsec:abstractgebra}) by \cd{Vector<X>::NumericType}.
It is required that it must be possible to add (or subtract) any two elements of the vector.
It is sometimes necessary to consider the characteristics of a zero-length vector, hence as well as indexing we also require a method
\begin{code}
    Vector<X>::zero_element() -> X
\end{code}

The operations supported by a vector $\mathbb{V}$ over some scalar type $\mathbb{X}$ are addition $\V + \V \fto \V$ and scalar multiplication $\X \cdot \V \fto \V$, implemented as \cd{operator+} and \cd{operator*}.
Further, in \Ariadne, all vectors are expressed in a canonical basis so support subscripting $\V[\N] \fto \X$ implemented as
\begin{code}
    Vector<X>::operator[](SizeType) -> ScalarType}
\end{code}
\Ariadne\ provides two basic vector norms, supremum (the default), and Euclidean norm, defined by
\[ \textstyle \|v\|_\infty = \sup\{ |v_i| \mid i=1,\ldots,n\}; \quad \|v\|_2 = \sqrt{\sum_{i=1}^{n} v_i^2} \]
and implemented as
\begin{code}
    sup_norm(Vector<X>) -> Positive<X>;
    two_norm(Vector<X>) -> Positive<ArithmeticalType<X>>;
\end{code}

The \cd{Matrix<X>} class provides dense matrices, and the \cd{SparseMatrix<X>} class sparse matrices. The standard arithmetical operations are provided, as are linear-equation solvers for solving $Ax=b$, notably
\cd{lu\_solve} using Gaussian eliminarion, and \cd{gs\_solve} using the Gauss-Seidel method, each with signature
\begin{code}
    solve(Matrix<X> A, Vector<X> b)
        -> Vector<ArithmeticalType<X>>
\end{code}

\subsection{Abstract algebra}
\label{subsec:abstractgebra}

An \emph{algebra} $\mathbb{A}$ over a field $\X$ (which we henceforth take to be the real numbers $\R$) is a type supporting addition, multiplication and scalar multiplication
\[ \A + \A \fto \A ; \qquad \A \times \A \fto \A; \qquad \X \cdot \A \fto \A \]
such that $\A$ is a vector space over $\X$, and multiplication satisfies $(c\cdot a)\times b = c\cdot(a\times b) = a\times (c\cdot b)$ and $a\times(b+c)=a\times b + a\times c$.
An algebra is \emph{associative} if multiplication is associative, and \emph{commutative} algebras if multiplication is commutative.
An algebra is \emph{unital} if it has a multiplicative identity~$1$; in this case we also have a scalar addition operation
\(  \X + \A \fto \A \)
defined by $c+a := c \cdot 1 + a$.
Note that most algebras should be thought of as representing \emph{scalar} quantities with some additional structure, rather than as \emph{vectors} over $\X$.

In \Ariadne, the property of being an algebra is defined by the C++ concepts \cd{AnAlgebra<A>} and \cd{AnAlgebraOver<A,X>}, where the former requires a type \cd{A::NumericType} giving the underlying field \cd{X} of \cd{A}.
Examples of algebras in \Ariadne\ include polynomials over one or many variables, denoted by the class \cd{Polynomial<I,X>} where \cd{I} is the \emph{index} type and \cd{X} the \emph{coefficient} type, symbolic \cd{RealExpression}s, continuous \cd{Function<P,Real(ARG)>} types, and \cd{Differential<X>} objects for automatic differentiation.
The numeric type \cd{X} may be a \cd{ValidatedNumber}, or a concrete number type such as for the \cd{ValidatedTaylorModel<F>} classes used as a core concrete function type, as detailed in Section~\ref{sec:functioncalculus}.

An \emph{elementary algebra} additionally supports elementary operations (except perhaps $\abs$), an \emph{analytic algebra} supports general analytic functions, and a \emph{continuous algebra} supports continuous operation.
Continuous algebras include the algebra of continuous functions $\X\to\R$, since given any $h:\R^n\to\R$, we can define $h(f_1,\ldots,f_\rndn) : x\mapsto h(f_1(x),\ldots,f_\rndn(x))$.
(Note that in the current implementation, \cd{Function<Real(ARG)>} only supports elementary operations.)
Similarly, \cd{RealExpression} can be defined as a pointwise algebra, but again, only elementary operations are currently supported.
Given an algebra \cd{A} over \cd{X}, the syntax \cd{apply(f,a)} applies \cd{f} to~\cd{a}.

\label{subsec:banachalgebra}

Elements of a \emph{Banach algebra} have a \emph{norm} $\|a\|$, expressed expressed \cd{a.norm()} or \cd{norm(a)}.
Further, it is often useful to be able to give a number $c$ such that $\|a-c\|$ is minimised; this number is the \cd{unit\_coefficient()}, denoted $\langle \A \rangle \fto \X$.
In order to support validated operations, we use the unit ball $\hat{e}=\{ a\in \A \mid \|a\|\leq 1\}$.

For elements of a unital Banach algebra, it is possible to compute arbitrary analytic functions using the power-series expansion.
The exponential $\hat{r}=\exp(x)$ can be expressed as
\[ \begin{gathered}
     c = \langle x \rangle; \quad
     y = x - c; \quad
     k = \max\{0,\lceil \log_2 \|y\| \rceil\}; \quad
     z = (1/2^k) \cdot y; \\[\jot]
    \hat{s} = {\textstyle\sum_{n=0}^{N-1}} z^n/n! + (\|z\|^N/N!) \hat{e}; \qquad
    \hat{r} = \exp(c) \cdot \hat{s}^{(2^k)}
\end{gathered} \]
Note that the power $\hat{s}^{2^k}$ can be computed by squaring $k$-times.

The \cd{TaylorModel<I,F,FE>} classes described in Section~\ref{subsec:polynomialmodelcalculus} form a Banach algebra, and analytic operations on them are implemented using generic Banach algebra algorithms.

\subsection{Differential algebra}
\label{subsec:differentialalgebra}

A \emph{differential} algebra over $n$ variables supports derivative operations $\partial_j$ for $j=1,\ldots,n$, which are an abstraction of partial derivatives.
These are linear operators which commute and satisfy the Leibnitz rule:
In a unital algebra, we have $\partial_i c = 0$ for a constant $c$.
An element $v_i$ is a \emph{variable} if $\partial_j v_i = \delta_{i,j}$.

Elements of a differential algebra can be used to implement automatic differentiation of functions.
\Ariadne\ provides a \cd{Differential<X>} class which models a quantity and its derivatives with respect to independent variables.
For efficiency reasons, rather than store the derivatives, we store the coefficients of the corresponding polynomial, and extract the derivatives using
\[ \displaystyle \frac{\partial^{|\alpha|} x^\alpha}{\partial x_1^{\alpha_1}\cdots\partial x_\rndn^{\alpha_\rndn}} =  \prod_{i=1}^{n}\alpha_i! \  . \]

Versions optimized for first and second order derivatives only, and for univariate functions, are also provided.
The \cd{FirstDifferential<X>} class only stores first derivatives, and implements the data
\[ (u;\,\p_1{u},\ldots,\p_\rndn{u})  \]
where each ``$\partial_{j}u$'' is a number representing the value of $\partial{u}/\partial{x_j}$ at a given point $x$.
The formulae for additon, multiplication and function application are then
\[ \begin{gathered}
  (u;\,\p_1{u},\ldots,\p_\rndn{u}) + (v;\,\p_1{v},\ldots,\p_\rndn{v}) = (u+v;\,\p_1{u}+\p_1{v},\ldots,\p_\rndn{u}+\p_\rndn{v}) ; \\
  (u;\,\p_1{u},\ldots,\p_\rndn{u}) \times (v;\,\p_1{v},\ldots,\p_\rndn{v}) = (u \times v;\;\p_1{u}\!\times\!v\!+\!u\!\times\!\p_1{v},\,\ldots,\,\p_\rndn{u}\!\times\!v\!+\!u\!\times\!\p_\rndn{v}) ; \\
    f(u;\,\p_1{u},\ldots,\p_\rndn{u}) = (f(u);\ f'\!(u)\,\p_1{u},\;\ldots,\;f'\!(u)\,\p_\rndn{u}) .
\end{gathered} \]

Differential algebras are \emph{graded}, meaning that any element can be written in the form
$ \textstyle a = \sum_{i=0}^{\infty} a_i$ where each $a_i$ has total degree $i$.
Hence constants $c\cdot 1$ lying in $\A_0$, and $a\in A_k$ if $\partial_i a \in A_{k-1}$ for all (any) $i$.
Multiplication in a graded algebra satisfies
\[ \textstyle \Bigl( \sum_{i=0}^{\infty} a_i\Bigr) \times \Bigl( \sum_{i=0}^{\infty} b_i \Bigr) = \sum_{i=0}^{\infty} \Bigl(\sum_{j=0}^{i} a_j b_{i-j}\Bigr) . \]
Since elements of grade $i$ do not depend on elements of grade $j$ for $j>i$, graded algebras are analytic.
This property is used to in generic code to apply the elementary operators to the \cd{Differential} classes.

\section{Function Calculus}
\label{sec:functioncalculus}

The main generic function types are \cd{EffectiveFunction} for exact scalar functions described symbolically, and \cd{ValidatedFunctionPatch} for interval functions on a box domain.

\subsection{Generic functions}
\label{subsec:genericfunctions}

The main generic function classes in \Ariadne\ are provided by the template \cd{Function<P,RES(ARGS...)>}, where \cd{P} is the information tag, \cd{RES} is the result type, and \cd{ARGS...} the argument types.
Currently, functions with \cd{EffectiveTag}, \cd{ValidatedTag} and \cd{ApproximateTag} information are supported, and the argument and result types can be \cd{Real} or \cd{Vector<Real>}, or synonymously \cd{RealScalar} and \cd{RealVector}.

A function with a \cd{Scalar} argument is \cd{Univariate}, and with a \cd{Vector} argument \cd{Multivariate}.
Hence \cd{ValidatedScalarMultivariateFunction} is a synonym for the template \cd{Function<ValidatedTag,RealScalar(RealVector)>}.
We also define partially-templated aliases, so \cd{ValidatedFunction<SIG>} is short for \cd{Function<ValidatedTag,SIG>} and \cd{ScalarMultivariateFunction<P>} for \cd{Function<P,RealScalar(RealVector)>}.

Functions may be constructed using the C++ ``named constructors'' idiom:
\begin{code}
    Function<P,RES(ARG)>::constant(...);
    Function<P,RES(ARG)>::coordinate(...);
\end{code}
or using named variables e.g.
\begin{code}
    x = RealVariable("x"); t=RealVariable("t");
    f = EffectiveScalarMultivariateFunction([x,t],x*exp(t));
\end{code}
Functions form an elementary algebra, so \cd{f1*f2} denotes the function $x\mapsto f_1(x) \times f_2(x)$, and  \cd{sin(f)} denotes the function $x\mapsto \sin(f(x))$.
Functions may be composed, with \cd{compose(f,g)} denoting $f\circ g$.
Other operations include \cd{derivative(f,k)} for $\partial{f}/\partial{x_k}$, where $k$ may be omitted if $f$ is univariate, \cd{join(f1,f2)} denoting $x\mapsto(f_1(x),f_2(x))$ and \cd{combine(f1,f2)} for $(x_1,x_2)\mapsto(f_1(x_1),f_2(x_2))$.

Extended evaluation operations are provided, so \cd{ApproximateFunction<Real(Real)>} can be evaluated not only on \cd{ApproximateNumber}, but also on \cd{FloatDPApproximation} and \cd{FloatMPApproximation}, and indeed, over any pointwise-algebra over \cd{ApproximateNumber}.
Similarly, \cd{ValidatedFunction<Real(Real)>} also supports evaluation on \cd{ValidatedNumber}, \cd{FloatBounds<PR>} and \cd{FloatBall<PR,PRE>}.

Derivatives can be computed by evaluating over \cd{Differential} objects as defined in Section~\ref{subsec:differentialalgebra} and polynomial approximations by evaluating over \cd{TaylorModel} objects (described in Section~\ref{subsec:polynomialmodelcalculus}).

Functions are assumed to be total, but may throw a \cd{DomainException} if they are passed an invalid argument, such as one causing a divide-by-zero.

\subsection{Function patches and models}
\label{subsec:functionpatchmodel}

Most function approximations are only valid over bounded domains.
The \Ariadne\ \cd{FunctionPatch} classes define partial functions over intervals $[a\!:\!b]$ of class \cd{IntervalDomainType} or coordinate-aligned boxes $D=\prod_{i=1}^{n}[a_i\!:\!b_i]$ of class \cd{BoxDomainType}.
The \cd{domain} method returns the domain of the function.
Concrete calculations are performed by \cd{FunctionModel} objects, which also specify concrete numerical precision \cd{PR} used for evaluation and error bounds \cd{PRE}.
Unlike \cd{Function} objects, \cd{FunctionModel}s are \emph{mutable}, and support in-place modification e.g. using \cd{operator*=}.

A \emph{uniform function model} is defined by a tuple $\langle D, g, e \rangle$ where $D$ is a \emph{domain}, $g:D\rightarrow\R$ is a \emph{validated approximation} and $e$ is an \emph{error bound}.
The tuple represents any function $f:D\fto\R$ such that
\[ \textstyle \dom(f)\supset D \text{ and } \sup_{x\in D} |f(x)-g(x)|\leq e . \]
Typically the function $g$ is a finite linear combination of basis functions $\phi_\alpha$, so
\[ \textstyle  g(x) = \sum_{\alpha \in \aleph} c_\alpha \phi_\alpha(x) \]
where $\aleph$ is a finite set of indices.
When these basis functions are products of univariate basis functions $\phi_k$, we have
\[ \textstyle  g(x) = \sum_{\alpha \in \aleph} c_\alpha \prod_{i=1}^{n}  \phi_{\alpha_i}(x_i) \]
where $\alpha\subset\N^n$ is a \emph{multi-index}.
In particular, for the monomial basis $\phi_k=x^k$ we have $x^\alpha:=\prod_{i=1}^{n} x_i^{\alpha_i}$.
The coefficients $c_\alpha$ are typically floating-point numbers of a fixed precision, either \cd{FloatDP} or \cd{FloatMP}.

A \emph{unit} function model is defined over the unit box $D=[-1\!:\!+1]^n$.
The main advantage of using a unit domain is that important basis univariate functions, including the monomials $x^i$ and Chebyshev functions, have maximum absolute value equal to $1$ over $[-1\!:\!+1]$, making it easy to determine which coefficients have a large impact on the function behaviour.

Function models over other box domains are defined in terms of unit function models via a \emph{scaling} function.
This is an affine bijection $s:[-1\!:\!+1]^n\rightarrow \prod_{i=1}^{n}[a_1\!:\!b_i]$ defined componentwise by $[s(z)]_i=s_i(z_i)$.
If $c_i=(a_i+b_i)/2$ and $r_i=(b_i-a_i)/2$, then $s_i(z_i)=r_iz_i+c_i$ and $s_i^{-1}(x_i)=(x_i-c_i)/r_i$.
\begin{figure}[h]
\centering
\includegraphics[scale=0.7]{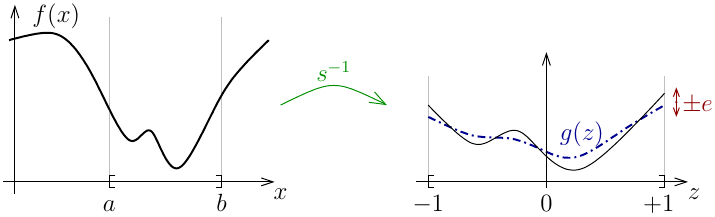}
\caption{A scaled function model.}
\end{figure}
A scaled function model is defined by a tuple $\hatf=\langle s,g,e\rangle$ where $s$ is a scaling function with codomain $D$ and $g$ defined on the unit domain.
The scaled function model $\hatf$ represents a function $f$ if $\dom(f)\supset D$ and
\[ \textstyle \|g\circ s^{-1}-f\|_{\infty,D} := \sup_{x\in D} | g(s^{-1}(x)) - f(x) | \leq e,  \]
or equivalently if
\[ \|g-f\circ s\|_{\infty,[-1:+1]^n}  \leq e,  \]

A validated function model $\hatf$ represents a set of functions $f$, and has operations preserving the inclusion property of Section~\ref{subsec:rigorousnumerics}.
Hence for a binary operation $\star$, we must have
\[ f_1 \in \hatf_1 \ \wedge\ f_2\in\hatf_2 \ \implies \  f_1 \star f_2 \in \hatf_1\; \hat{\star} \; \hatf_2 . \]

\subsection{Polynomial models}
\label{subsec:polynomialmodelcalculus}

The main class of validated function model in \Ariadne\ are the \cd{TaylorFunctionModel}s, based on the Taylor models of~\citet{MakinoBerz2003}.
Since the only existing package implementing Taylor models when \Ariadne\ was under development, COSY Infinity~\citep{MakinoBerz2006cosy}, is not open-source, an implementation within \Ariadne\ itself was developed.
Unlike the original Taylor models, where an arbitrary interval $I$ is used as the remainder term, here we only give an error bound $e$, corresponding to an interval remainder $I=[-e\!:\!+e]$.

In \Ariadne, a \emph{scaled polynomial model} or \emph{scaled Taylor model} is a scaled polynomial approximation on a box, defined by a tuple $\langle s,p,e \rangle$ where
\begin{enumerate}[(i)]
 \item each $s_i:[-1\!:\!+1]\fto[a_i\!:\!b_i]$ is a scaling function $z_i\mapsto r_iz_i+c_i$.
 \item $p:[-1\!:\!+1]^n\fto \R$ is a polynomial $z\mapsto\sum_{\alpha} c_\alpha z^\alpha$ with each $c_\alpha\in\F$.
 \item $e\in\F_\mathrm{err}^+$ is an error bound.
\end{enumerate}
Here $\mathbb{F}$ and $\mathbb{F}_\mathrm{err}$ are concrete numerical types, such as \cd{FloatDP} or \cd{FloatMP}.
The \emph{domain} of the model is $D=\prod_{i=1}^{n}[a_i\!:\!b_i]$.

A scaled polynomial model $\langle s,p,e\rangle$ represents $f:\R^n\fto\R$ if
\begin{equation*} \textstyle \sup_{x\in D} \bigl| f(x) - p\circ s^{-1}(x) \bigr| \leq e . \end{equation*}
In particular, we can only represent functions over a bounded domain.

Unit polynomial models $p\pm e$ are defined by \cd{TaylorModel<ValidatedTag,F,FE>}, where \cd{F} is the type used for the coefficients of the polynomial and \cd{FE} the type used for the error bound,
The current implementation uses a sparse representation of the polynomial, with separate lists for the multi-indices $\alpha \in \N^n$ and the coefficients $c \in \F$.
These are ordered by \emph{reverse lexicographic order}, which makes inserting a constant term fast, and allows efficient evaluation using Horner's rule.
Alternative representations include a \emph{dense} format, with only coefficients being stored, which may be useful for low-dimensional polynomials, and a representation using sparse storage for the multi-indices themselves, which may be useful for very high-dimensional polynomials.

Different methods are provided to evaluate and compose polynomial models, including direct evaluation and evaluation based on Horner's rule.
For narrow interval vectors, a standard Horner evaluation is sufficient, which for univariate polynomials is
\begin{equation*} p(x) = c_0+x(c_1+x(c_2+\cdots x(c_{n-1}(x+c_{n}x)))) . \end{equation*}
Evaluation can be performed using function call syntax via \cd{operator()}, or by the free function \cd{evaluate}.

Since the problem of finding tight bounds for the range of a polynomial is known to be NP-complete, heuristics are used to balance the trade-off between speed and accuracy of the result.
The current implementation of the \cd{range} uses a simple over-approximation
\begin{equation*} \range \bigl( p \pm e\bigr) \subset c_0 \pm \bigl( \sumup_{\alpha\neq0} |{c_\alpha}| \,+_\rndu\, e ) . \end{equation*}
Similarly, the \cd{norm} function computes an over-approximation to the uniform norm by
\begin{equation*} \| p \pm e \|_{\infty,[-1:+1]^n} \leq  \sumup_{\alpha} |{c_\alpha}| \,+_\rndu\, e . \end{equation*}

The \cd{refines} function tests whether one polynomial model refines another.
Note that $p_1\pm e_1$ is a refinement of $p_2\pm e_2$ if $\|p_1-p_2\|+e_1 \leq e_2$.
This can be checked using the over-approximation to the norm function, yielding sufficient condition:
\begin{equation*} \textstyle \sumup_\alpha |c_{1,\alpha} - c_{2,\alpha}|_\rndu +_\rndu e_1 \leq e_2 . \end{equation*}

To avoid growth of the number of coefficients, we can \emph{sweep} small coefficients into the uniform error.
Removing the term $c_\alpha x^\alpha$ introduces an error of size $|c_\alpha|$, so we can show
 \begin{equation*} \textstyle \bigl(\sum_\alpha c_\alpha x^\alpha + \sum_\beta c_\beta x^\beta\bigr) \pm e \prec \sum_\alpha c_\alpha x^\alpha \pm (\sumup_\beta|c_\beta| \, +_\rndu e) . \end{equation*}
Here, we make critical use of the fact that the domain of $p$ is $[-1\!:\!+1]^n$.
More sophisticated sweeping schemes are possible;
finding good sweeping schemes is a crucial ingredient in the \emph{efficiency} of polynomial model arithmetic.
Various \cd{Sweeper} classes allow the accuracy versus efficiency tradeoff of the polynomial to be controlled.

Ordinary arithmetical operations can be performed on polynomial models.
Care must be taken to ensure roundoff errors in floating-point arithmetic are handled correctly.
For example, addition is performed using
\begin{equation*} \begin{aligned}
\textstyle (p_1\pm e_1)+(p_2\pm e_2) =& \textstyle \sum_{\alpha} (c_{1,\alpha} +_\rndn c_{2,\alpha}) x^\alpha \\
    & \qquad \pm   \sumup_\alpha  e_{u}(c_{1,\alpha},+,c_{2,\alpha})  +_\rndu (e_1 +_\rndu e_2).
\end{aligned} \end{equation*}
where $e_\rndu(x,+,y)$ is an upper bound of $|(x+y) - (x+_\rndn y)|$, as defined in Section~\ref{subsec:numbers}.
Currently the scheme $e_\rndu(x,+,y) := \bigl( (x +_\rndu y) -_\rndu (x+_\rndd y) \bigr) \div_\rndu 2$ is used, but a more faster, though less accurate, alternative would be to use $\frac{1}{2}\ulp(x +_\rndn y)$.

To define multiplication, we first need to consider the error bound.
Since in the uniform norm we have
\[  \| f_1 \times f_2 - p_1 \times p_2\| \leq \|p_1\|\!\cdot\!\|f_2-p_2\| + \|f_1-p_1\|\!\cdot\!\|p_2\| + \|f_1-p_1\|\!\cdot\!\|f_2-p_2\|  \]
the algorithm for computing $(p_1\pm e_1)\times(p_2\pm e_2)$ reduces to computing the product $p_1\times p_2$.
For this the current algorithm uses term-by-term monomial multiplication
\begin{equation*} \begin{aligned}
\textstyle (\sum_\alpha c_\alpha x^\alpha) \times c_\beta x^\beta = & \textstyle \sum_{\alpha} (c_\alpha \times_\rndn c_\beta) x^{\alpha+\beta}  \\
    & \qquad \textstyle  \pm   \sumup_\alpha  e_{u}(c_{\alpha},\times,c_{\beta})
\end{aligned} \end{equation*}
and adds the resulting polynomials using the addition algorithm.

Composition in \Ariadne\ is provided by the \cd{compose} function.
Separate implementations are available for precomposing with a polynomial model or a general function.

If $\hat{g}_i=\langle r,p_i,e_i\rangle $ are scaled polynomial models and $\hatf=\langle s,q,d \rangle$ is a scaled polynomial model with $\dom(\hatf)\supset\range(\hat{g})$, then the composition $\hatf\circ \hat{g}$ can be computed by applying the polynomial expression $q\pm d$ to each $s_i^{-1}\circ (p_i\pm e_i)$.
We obtain
\begin{equation*} \hatf\circ \hat{g} = q\bigl(s_1^{-1}\circ(p_1\pm e_1),\ldots,s_\rndn^{-1}\circ(p_\rndn\pm e_\rndn)\bigr)\pm d . \end{equation*}
The resulting computation uses only addition and multiplication, so can be implemented using basic function model arithmetic.
The polynomial $q$ is evaluated using Horner's to improve the efficiency of the subsequent evaluations.

If $f:\R\rightarrow\R$ is an analytic function,  and $\hat{g}$ is a polynomial model with $\range(\hat{g}) \subset c\pm r$, then by Taylor's theorem with remainder,
\begin{equation*} f\circ\hat{g} =+ \sum_{i=0}^{n} f^{(i)}(c) \hat{g}^i \pm \; | f^{(n)}([a\!:\!b])-f^{(n)}(c)|\,r^n . \end{equation*}

If $f$ is an elementary function, then $f\circ \hat{g}$ is computed by applying the component functions in order.
Alternatively, a polynomial model approximation $\hatf$ can be constructed and applied to $\hat{g}$.
It is unclear which approach yields best results in practise.

The \cd{antidifferentiate} method computes an indefinite integral with respect to a variable:
  \begin{equation*} \int p\circ s^{-1} \,\mathrm{d}x_j \circ s = \smfrac{b_j-a_j}{2} \Bigl( \sum_\alpha \tfrac{1}{\alpha_j+1} \,c_\alpha\,  x^{\alpha+\epsilon_j} \pm e \Bigr). \end{equation*}
If $e=0$, then the \cd{derivative} function allows a polynomial model to be differentiated term-by-term:
  \begin{equation*} \smfrac{\mathrm{d}(p\circ s^{-1})}{\mathrm{d}{x}_j} \circ s = \smfrac{2}{b_j-a_j} \sum_\alpha \alpha_j c_\alpha x^{\alpha-\epsilon_j} . \end{equation*}
If $e>0$, then there are non-differentiable functions captured by the polynomial model.
However, the derivative of the \emph{midpoint} model $(s,p,0)$ can often be used in calculations, notably to solve algebraic equations.

The \cd{split} functions split the domain $D$ into subdomains $D_1, \ldots, D_r$ and replace the polynomial model $p\circ s^{-1}\pm e$ with the corresponding \emph{restrictions} $p_j\circ s_j^{-1}\pm e_j$ where $s_j:D_j\to[-1\!:\!+1]^n$. The restriction from domain $D$ to $D_i$ is performed by precomposing $p$ with the rescaling function $s^{-1}\circ s_j$, so $p_j = p \circ (s^{-1}\circ s_j)$.

\Ariadne\ provides two main approaches to constructing a polynomial model for a function $f$.
If $f$ defined in terms of elementary functions, we can compute the composition
\( \hatf = f\circ\widehat{\id} \) to obtain a polynomial model for $f$.

Taylor polynomial models can also be computed from the Taylor series with remainder term.
In one-dimension over the interval $c\pm r$ we have (using exact arithmetic).
\begin{equation*} \textstyle \hatf(x) = \sum_{i=0}^{n} (f^{(i)}(c)/r^i) x^i \; \pm \, |f^{(n)}([a\!:\!b])-f^{(n)}(c)|/r^n . \end{equation*}
When implemented using rounded arithmetic, the values $f^{(k)}(c)$ are computed as intervals, and the accumulated roundoff errors are swept into the uniform error term.

\subsection{Affine models}
\label{subsec:affinemodelcalculus}

An \cd{AffineModel} over $[-1\!:\!+1]^n$ is a function of the form
\[ \textstyle \hatf(x) = b + \sum_{i=1}^{n} a_i x_i \pm e . \]
The product of two affine models cannot be evaluated exactly, even in exact arithmetic, since quadratic terms need to be discarded.
We obtain
\begin{multline*} \bigl( b_1 + {\textstyle\sum_{i=1}^{n}} a_{1,i} x_i \pm e_1\bigr) \times \bigl( b_2 + {\textstyle\sum_{i=1}^{n}} a_{2,i} x_i \pm e_2 \bigr)
    \\ \mbox{}\hspace{-3em} = b_1b_2 + {\textstyle\sum_{i=1}^{n}} (b_1a_{2,i}+b_2a_{1,i}) x_i
    \\ \pm \tsum|a_{1,i}|\cdot \tsum|a_{2,i}| + \bigl(|b_1|+\tsum|a_{1,i}|\bigr)e_2 + \bigl(|b_2|+\tsum|a_{2,i}|\bigr)e_1 + e_1 e_2 .
\end{multline*}

Affine models are used in \Ariadne\ in conjunction with specialised methods for solving linear differential equations and linear constrained optimisation problems; for more general operations polynomial models are preferred.

\section{Solving Algebraic and Differential Equations}
\label{sec:solvingequations}

Having provided the basic function calculus and forward operations of evaluation and composition, we turn to more complicated problems of solving equations.
In \Ariadne, these operations are implemented by \emph{solver} classes.
The advantage of using solver classes is that multiple solvers can be provided for the same kind of problem, enabling the solving policy to be specified at run-time.
Further, solvers can encapsulate details of their accuracy parameters, which means that the calling syntax need only specify the problem variables, simplifying use and improving encapsulation.
Additionally, (sensible) default parameters can be provided so that users do not need to know the details of the internal workings of the class.

\subsection{Algebraic equations}
\label{subsec:algebraicequations}

Solvers for a simple algebraic equation $f(y)=0$ or a parameterised algebraic equation $f(x,y)=0$ with solution $y=h(x)$ are simply called \cd{Solver}.
\Ariadne\ implements an \cd{IntervalNewtonSolver} based on the interval Newton operator~\citep{Moore1966}, and a \cd{KrawcykzSolver} based on the related Krawcykz operator~\cite{Krawczyk1969}, which is more reliable but slower.

The method
\begin{code}
    SolverInterface::
        solve(ValidatedVectorMultivariateFunction f,
              ExactBoxType D)
            -> Vector<ValidatedNumber>;
\end{code}
attempts to find a solution of the equation $f(x)=0$ in box $D$, throwing a \cd{SolverException} if no solution is found.
The method delegates to an abstract method
\begin{code}
    SolverInterface::
       step(ValidatedVectorMultivariateFunction f,
            Vector<ValidatedNumber> x)
           -> Vector<ValidatedNumber>;
\end{code}
which performs a single step of the algorithm.

\newcommand{\aprx}[1]{\check{#1}}

A single step of the solver computes an operator $S(f,\hat{y})$ with the property that any solution to $f(y)=0$ in $\hat{y}$ also lies in $S(f,\hat{y})$, and further, if $S(f,\hat{y})\subset\hat{y}$, then the equation $f(y)=0$ has a unique solution in $\hat{y}$.
Note that if $S(f,\hat{y})\cap\hat{y}=\emptyset$, then there are no solutions in $\hat{y}$.
The step may fail, in which case no information about the solutions can be deduced.
Tight bounds to the solution are found by the iteration
\[ \hat{y}_{n+1} = S(f,\hat{y}_\rndn) \cap \hat{y}_\rndn . \]

A single step of the interval Newton solver is given by
\begin{equation*} \mathrm{N}_\mathrm{ivl}(f,\hat{y},\aprx{y}) =  \aprx{y} - [\jacob{f}(\hat{y})]^{-1} f(\aprx{y}), \end{equation*}
where $\aprx{y}$ is an arbitrary point in $\hat{y}$, typically chosen to be the midpoint.
If $\jacob{f}(\hat{y})$ contains a singular matrix, then the interval Newton method fails and provides no information.
A single step of the Krawczyk solver is given by
\begin{equation*} \mathrm{Kr}(f,\hat{y},\aprx{y},\aprx{J}) =  \aprx{y} - \aprx{J}^{-1}f(\aprx{y}) + (I-\aprx{J}^{-1}\jacob{f}(\hat{y}))(f(\hat{y})-f(\aprx{y})) \end{equation*}
where $\aprx{J}^{-1}$ is an arbitrary matrix, typically an approximation to the inverse Jacobian $Df(\aprx{y})^{-1}$.

To solve the parameterised equation
\( f(x,h(x))=0 \)
where $f:\R^n\times\R^m\to\R^m$ and the solution $h$ is a partial function $h:D\subset\R^n\to\R^m$, we use a parameterised interval Newton operator
\begin{equation*} \mathrm{N}_\mathrm{ivl}(f,\hat{h},\aprx{h}) = x \mapsto \aprx{h}(x) - [\jacob_2{f}(x,\hat{h}(x))]^{-1} f(x,\aprx{h}(x))  \end{equation*}
similar to~\citep{BerzHoefkens2001}.
If $f:\R^n\times\R\fto\R$, then
\begin{equation*} \mathrm{N}_\mathrm{ivl}(f,\hat{h},\aprx{h})(x) =  \aprx{h}(x) - f(x,\aprx{h}(x))/f_{,y}(x,\hat{h}(x)) . \end{equation*}

\begin{rem}
Solutions of linear algebraic (matrix) equations are also possible, but are defined by functions such as \cd{lu\_solve} rather than solver classes.
Solver classes for matrix equations will be provided in a future release.
\end{rem}

\subsection{Differential equations}
\label{subsec:differentialequations}

An \emph{integrator} is a evaluator for computing the flow of a differential equation.
\Ariadne\ currently provides three integrators, an \cd{AffineIntegrator} for computing the flow of an affine system $\dot{x}=Ax+b$, a \cd{PicardIntegrator}, which uses Picard's iteration,
and a \cd{TaylorIntegrator}, which computes the flow using a Taylor series expansion.
In practise, the Taylor integrator outperforms the Picard integrator, since it generates sharper error bounds after fewer iterations.

The \emph{flow} $\phi(x,t,a)$ of the parameterised differential equation $\dot{y}=f(y,a)$ is defined by
 \begin{equation*} \dot{\phi}(x,t,a) = f(\phi(x,t,a),a); \quad \phi(x,0,a) = x . \end{equation*}
If there are no parameters, we drop the variable $a$, obtaining $\dot{\phi}(x,t)=f(\phi(x,t))$.
In this case, a time step for the flow over the domain $T=[0,h]$ can be computed by
\begin{code}
    IntegratorInterface::
      flow_step(ValidatedVectorMultivariateFunction f,
                BoxDomainType X0,
                IntervalDomainType T)
          -> ValidatedVectorMultivariateFunctionPatch;
\end{code}
A general approach involves first finding a \emph{bound} for the flow, which is performed by a \cd{Bounder} class.
A bound for the flow starting in a set $D$ with time step $h$ is any (convex) set $B$ such that
\[ \forall t\in [0,h], \ \forall x\in D, \ \phi(x,t) \in B . \]
It can be shown that a convex set $B\supset D$ is a bound if $B \supset D+h f(B)$, and further, if $B$ is a bound, then so is
\[ B' := D + [0,h] \convex(f(B)) \subset B, \]
so bounds can be iteratively refined.
To find an initial bound, note that for any neighbourhood $B$ of $D$, there exists $h>0$ such that $B$ is a bound for the time step $h$.
As a heuristic to find a bound, we take $B=\hatD+[0,2h]\convex(f(\hatD))$, where $\hatD$ is given by $c+2(D-c)$ where $c$ is the midpoint of $D$.
If a bound $B$ is known, then \cd{flow\_step} can be called with an extra parameter \cd{BoxDomainType B}.

A simple method to compute the solution is to use Picard's iteration
\begin{equation*} \textstyle \phi_{n+1}(x,t,a) = x + \int_0^t f(\phi_\rndn(x_0,\tau,a),a)\,d\tau , \end{equation*}
which is a contraction for $t\in[0,h]$ if $h<1/L$ where $L$ is the Lipschitz constant of $f$ with respect to the state $y$.
Since the only operations used are composition and antidifferentiation, which are computable on function models, we can apply Picard's iteration directly. If $\hat{\phi}_{n+1}$ refines $\hat{\phi}_\rndn$, then $\hat{\phi}_{n+1}$ is a function model for the flow, as is any further iteration.
Further, if $\hat{\phi}_0(x,t,a)$ is the constant box $B$, then $\hat{\phi}_1$ refines $\hat{\phi}_0$, automatically yielding a solution.

An alternative method is to compute the Taylor series of the flow and estimate the error bounds.
For $y=\phi(x,t)$ satisfying $\dot{y}=f(y)$, we find for any function $g(y)$ that
\[ \textstyle \frac{d}{dt} g(\phi(x,t)) = \nabla{g}(\phi(x,t)) \cdot f(\phi(x,t)) \]
This allows us to compute time derivatives of the flow $\phi(x,t)$ by taking $g_k(\phi(x,t))=\frac{d}{dt}g_{k-1}(\phi(x,t))=\frac{d^k}{dt^k}\phi(x,t)$.
Time derivatives of the spacial derivatives $\partial^{|\alpha|} \phi(x,t) / \partial x^\alpha$ can be computed similarly.
The \cd{TaylorIntegrator} uses the \cd{Differential<X>} class to compute all partial derivatives~$\partial^{|\alpha|+k}\phi(x,t)/\partial x^\alpha\partial t^k$ together.

Let $x_c$ be the midpoint of $D$ and $B$ be a bound for $\phi(D,[t_0,t_1])$.
Then
\[ \begin{aligned}
\textstyle  \hat{\phi}(x,t)
                  =&\textstyle \sum_{|\alpha|\leq m \wedge k\leq n} c_{\alpha,k} (x-x_c)^\alpha (t-t_0)^k \\&\qquad  \textstyle \pm \sum_{\substack{|\alpha|\leq m \wedge k\leq n\\|\alpha|=m\vee k=n}} |C_{\alpha,k}-c_{\alpha,k}|\, |x-x_c|^\alpha\, |t-t_0|^k
\end{aligned}\]
where
\[ c_{\alpha,k} = \left.\frac{\partial\phi}{\partial x^\alpha \partial t^k}\right|_{x_c,t_0} \quad\text{and}\quad  C_{\alpha,k} = \left.  \frac{\partial\phi}{\partial x^\alpha \partial t^k}\right|_{B\times[t_0,t_1]} . \]
In practise, this Taylor method yields better results than the Picard method.

\subsection{Hybrid systems}
\label{subsec:hybridsystems}

A \emph{hybrid system} is a dynamic system where the state $x$ evolves continuously via $\dot{x}=f(x)$ until some \emph{guard condition} $g(x)\geq0$ is satisfied, at which time the state instantaneously jumps to another state via a \emph{reset} $x'=r(x)$.

Assuming an set $X_0$ of possible starting states, the dependence of the current state on the initial state $x$ and the current time $t$ is described by a function $\psi(x,t)$.
During continuous evolution, we the solutions $\psi$ are found by solving the differential equation $\dot{\psi}(x,t)=f(\psi(x,t))$, and during a discrete jump the solution is updated by composition $\psi'(x,t) = [r\circ \psi](x,t)$.
To determine the crossing times $\gamma$, we need to solve the parameterised algebraic equation $g(\psi(x,t))=0$ for $t=\gamma(x)$.
Using the interval Newton iteration
 \begin{equation*} \hat{\gamma}_{n+1}(x) = \aprx{\gamma}_\rndn(x) - \dfrac{g(\psi(x,\aprx{\gamma}_\rndn(x))}{(\nabla g\cdot f)(\psi(x,\hat{\gamma}_\rndn(x)))} , \end{equation*}
where $\hat{\gamma}$ is a validated function model for $\gamma$ containing $\aprx{\gamma}$.

We therefore see that the function calculus we have described provide exactly the building blocks needed to rigorously compute the evolution of a hybrid system.
For more details, see~\citep{CollinsBresolinGerettiVilla2012ADHS} and the website~\citep{ariadne}.

\lstset{language=Python}
\lstset{numberstyle=\color{blue}}

\section{Examples}
\label{sec:examples}

\newcommand{\ext}{\mathrm{ext}}

We now give examples of the use of \Ariadne's function calculus for solving algebraic and differential equations using the Python interface.
We shall use as a running example the \emph{Fitzhugh-Nagumo} system~\citep{Fitzhugh1961BIOPHYS,NagumoArimotoYoshizawa1962IRE}, which is defined by the equations
\[ \dot{v}=v-v^3/3-w+R\,I_\ext, \qquad \dot{w} = (v+a-bw)/\tau \]
where $\tau$ is the \emph{time-scale} of the slow variable $w$, and $I_\ext$ is an external forcing current.
Unless otherwise mentioned, we take parameter values
\[ \alpha=0.7, \quad \beta=2.0, \quad \tau=12.5, \quad R=0.1, \quad I_\ext=3.5 . \]

\subsection{Fixed points}

Consider the problem of finding the fixed-points of the Fitzhugh-Nagumo, which are given by solving the equations $\dot{v}=\dot{w}=0$.

To solve this in \Ariadne, we first define the Fitzhugh-Nagumo system itself:
\begin{code}
    a=Decimal(0.7); b=Decimal(2.0)
    tau=Decimal(12.5)
    R=Decimal(0.1); Iext=Decimal(3.5)
    v=RealVariable("v"); w=RealVariable("w")
    f=Function([v,w],[v-(v*v*v)/3-w+R*Iext,(v+a-b*w)/tau])
\end{code}
We need to define a bounded domain in which to search for the fixed-points:
\begin{code}
    domain=BoxDomainType([{-2:3},{-1:2}])
\end{code}
We then construct the solver class; here we use the interval Newton solver:
\begin{code}
    tolerance=1e-12; max_steps=32
    solver=IntervalNewtonSolver(tolerance,max_steps)
\end{code}
Finally, we use the \cd{solve\_all} method to compute all the fixed-points.
\begin{code}
    fixed_points=solver.solve_all(g,domain)
    print("fixed_points:",fixed_points)
\end{code}
The result is
\begin{smallcode}
   fixed_points:
       [ [-1.2247448713915[9:8],-0.26237243569579[5:4]],
         [-0.0000000000000[-139:110],0.35000000000000[-8:6]],
         [1.224744871391[48:70],0.962372435695[776:813]] ]
\end{smallcode}

Given a fixed-point, we can consider the functional dependence on the parameters.
Suppose we vary $I_\ext$ over the range $[3.0\!:\!4.0]$, and look for the fixed-points near $(1.22,0.96)$

We make $I_\ext$ a variable and redefine $f$.
\begin{code}
    Iext=RealVariable("Iext")
    f=Function([Iext,v,w],[v-(v*v*v)/3-w+R*Iext,(v+a-b*w)/tau])
\end{code}
We then define the domain of $I_\ext$ and the range of $(v,w)$ to consider:
\begin{code}
    Idom=BoxDomainType([[dy_(3.0),dy_(4.0)]])
    vw_rng=BoxDomainType([[dy_(1.0),dy_(1.5)],
                         [dy_(0.75),dy_(1.25)]])
\end{code}
Finally, we use the \cd{implicit} method to compute the parametrised fixed-point $(v,w)$ as a function of $I_\ext$:
\begin{code}
    parametrised_fixed_point = solver.implicit(f,Idom,vw_rng)
    print("parametrised_fixed_point:",paramatrised_fixed_point)
\end{code}
The result is
\begin{smallcode}
    parametrised_fixed_point:
    VectorScaledFunctionPatch(
      dom=[{3.0:4.0}],
      rng=[{1.1712915:1.2720745},{0.93564590:0.98603711}],
      [ { -0.00007341*x0^6 +0.001752*x0^5 -0.01789*x0^4
          +0.1014*x0^3 -0.3482*x0^2 +0.7955*x0
          +0.2577  +/-0.00000578},
        { -0.00002506*x0^6 +0.0006315*x0^5 -0.006803*x0^4
          +0.04071*x0^3 -0.1479*x0^2 +0.3610*x0
          +0.5003  +/-0.00000301} ]  )
\end{smallcode}
Here, the result is displayed as a non-centred polynomial, with $x_0$ being the variable $I_\ext$.

Note that if we were to use a wider initial range such as $[0.75\!:\!1.75]\times[0.5\!:\!1.5]$ then the algorithm fails, throwing an \cd{UnknownSolutionException}.

\subsection{Differential equations}

To solve the Fitzhugh-Nagumo system, we use an \cd{Integrator} class.
We first define the point and a desired time interval:
\begin{code}
    IntervalDomainType(1.25,1.5)
    vw0=BoxDomainType([ [0,0], [0,0] ])
    h=Dyadic(1.0)
    tolerance=1e-3
    integrator = TaylorPicardIntegrator(tolerance)
    flow_step = integrator.flow_step(f,vw0,suggest(h))
    print("flow_step:",flow_step)
\end{code}
Note that currently the starting point must be represented as a singleton box.
The result is:
\begin{smallcode}
    flow_step:
    VectorScaledFunctionPatch(
      dom=[{0.0:0.0},{0.0:0.0},{0.0:0.25}],
      rng=[{-0.000365602:0.097590134},{-0.000080380:0.014675381}],
      [ { 0.1585*x2^2 +0.3493*x2  +/-0.000366},
        { 0.009520*x2^2 +0.05600*x2  +/-0.0000804} ]  )
\end{smallcode}
Note that the time range only goes up to $0.25$.

If we consider the flow of all points starting in the box $[0\!:\!1]\times[0\!:\!1]$ using
\begin{code}
    vw0=BoxDomainType([ [0,1], [0,1] ])
    flow_step = integrator.flow_step(f,vw0,suggest(h))
    print("flow_step:",flow_step)
\end{code}
we find
\begin{smallcode}
    flow_step:
    VectorScaledFunctionPatch(
      dom=[{0.0:1.0},{0.0:1.0},{0.0:0.09375}],
      rng=[{-0.10411338:1.1214595},{-0.0041834586:1.0056835}],
      [ { 0.5000*x0^2*x1*x2^2 -0.4200*x1*x2^2 -0.9667*x0^2*x2^2
              +0.7167*x0*x2^2 +0.1385*x2^2 -1.000*x1*x2
              -0.3451*x0^3*x2 +0.01758*x0^2*x2 +0.9953*x0*x2
              +0.3494*x2 +1.0000*x0  +/-0.00161},
        { -0.1600*x1*x2 +0.08000*x0*x2 +0.05600*x2
              +1.000*x1  +/-0.000434} ]  )
\end{smallcode}
Here, an even smaller step-size of $0.09375$ was used.

To use the \cd{TaylorSeriesIntegrator}, a maximal order for the power series must also be given:
\begin{code}
    order=8
    integrator = TaylorSeriesIntegrator(tolerance,order)
    flow_step = integrator.flow_step(f,vw0,h)
    print("flow_step:",flow_step)
\end{code}
\begin{smallcode}
    flow_step:
    VectorScaledFunctionPatch(
      dom=[{0.0:1.0},{0.0:1.0},{0.0:0.09375}],
      rng=[{-0.10484287:1.1221622},{-0.0039900682:1.0055644}],
      [ { ... +1.0000*x0 +0.0000005001+/-0.00138},
        { ... +1.000*x1 -4.337e-19*x0 +2.103e-17+/-0.0000155} ]  )
\end{smallcode}
To compute more than a single time-step, an \cd{Evolver} class must be used.

\section{Extensions}
\label{sec:extensions}

We now briefly mention extensions to the basic function calculus of \Ariadne\ whose implementation is in progress.

\subsection{Alternative bases}

Work is in progress on using other bases than the monomial basis, as described in Section~\ref{subsec:functionpatchmodel}, notably the Chebyshev basis.
The underlying representation is the same,
\[ \textstyle p(x) = \sum_{\alpha} c_\alpha  \prod_{i=1}^{n} \phi_{\alpha_i}(x_i) \]
but different algorithms are needed for multiplication and evaluation.
The Chebyshev basis polynomials $T_k$ have products $T_j \cdot T_k = \tfrac{1}{2}(T_{|j-k|}+T_{j+k})$ so multiplication is harder than for the monomial basis.
However, they satisfy $\sup_{z\in[-1:+1]}T_k(z)=1$ and are orthogonal, so tighter bounds on the range can be produced than using the monomial basis.

Other classes of approximating function possible, such as Bernstein bases, Fourier series, neural networks, convolutions.

\subsection{Differentiable functions}

A differentiable ($C^1$) function calculus includes uniform errors on both zeroth and first derivatives.
We represent univariate $C^1$ function over a unit domain, $f:[-1\!:\!+1]\fto\R$ by a polynomial $p$, and provide error bounds
\[ \begin{gathered} {\textstyle\sup_{x\in[-1:+1]}}|f(x)-p(x)|\leq \delta_0,  \quad {\textstyle\sup_{x\in[-1:+1]}}|f'(x)-p'(x)|\leq \delta_1, \\ |f(0)-p(0)|\leq \tilde{\delta}_0 . \end{gathered} \]
In higher dimensions, we provide uniform bounds for each partial derivative.

Note that it is useful to provide both a uniform error and a punctual error at the midpoint of the domain.
The uniform error $\delta_0$ is bounded by $ \delta_0 \leq \tilde{\delta}_0 + \delta_1 $, but can usually be bounded more tightly by direct estimates.

The error bounds for addition are simply added, but for multiplication we use:
\[ \begin{aligned}
    \|f\cdot g - p\cdot q\| &\leq \|f'-p'\|\!\cdot\!\|q\| + \|p'\|\!\cdot\!\|g-q\| + \|f'-p'\| \!\cdot\! \|g-q\| +  \\
                    & \qquad + \|f-p\| \!\cdot\! \|q'\| + \|p\| \!\cdot\! \|g'-q'\| + \|f-p\| \!\cdot\! \|g'-q'\| .
\end{aligned}  \]
where $\|\cdot\|$ is the uniform norm over the domain $[-1\!+\!1]$.

\subsection{Multivalued functions}

Set-valued functions play an important role in defining nondeterministic dynamical systems.
Support for these functions is under development, with compact-valued functions being the most important.
As well as generic classes for set valued functions, concrete realisatations are also provided, such as multivalued functions of the form
\[ F(x) := \{ f(x,p) \mid p\in D \mid g(x,p) \in C \} . \]
Here, $p$ parametrised the set $F(x)$, with $D$ providing bounds for $p$ and $g(x,p)\in C$ a constraint.

\subsection{Measurable functions}

Since measurable functions are only defined up to equivalence almost everywhere, they cannot be computably evaluated.
Instead, a general representation is given in terms of preimages of open sets, with $f^{-1}(V)=U_\infty$ where $U_\infty$ is a \emph{lower-measurable set}~\citep{Collins2020ARXIV}:
\[ \textstyle U_\infty= \bigcap_{i=0}^{\infty}U_i \text{ where each $U_i$ is open, and } \mu(U_i\setminus U_j) \geq 2^{-j} \text{ if } i<j . \]
For measurable functions taking values in a separable complete metric space, this definition is equivalent to taking completions of continuous or piecewise-constant functions under the Fan metric
\( d(f_1,f_2) = \sup\!\big\{ \varepsilon\in\Q^+ \mid \ \mu\big(\{x \mid d(f_1(x),f_2(x))>\varepsilon\}\big) > \varepsilon\big\} . \)
Similarly, spaces of $p$-integrable functions are defined as completions under the $p$-norms.

\subsection{Weierstrass approximation}

By the Weierstrass approximation theorem, any continuous function $f:D\to\R$ over a bounded domain $D\subset\R^n$ can be uniformly approximated by polynomials.
For a univariate function on the unit domain, $f:[-1\!:\!+1]\to\R$, we can approximate by a unit polynomial model $\hatf=p\pm e$.
For a rigorous computation, we need to use only information provided by interval evaluation $\ivl{f}$.

\section{Conclusions}
\label{sec:conclusions}

\Ariadne\ is a software tool implementing a rigorous general-purpose calculus for working with functions over real variables.
The calculus includes support for interval arithmetic, linear algebra, automatic differentiation, function models with evaluation and composition, and solution of algebraic and differential equations.
Due to the modular structure, additional function types can be introduced, and different implementations of the algorithms provided.
We have shown examples of the rigorous solution of algebraic and differential equations, and seen how the functionality is sufficient to rigorously compute the evolution of nonlinear hybrid dynamic systems.

Work in the immediate future will focus on further improvements to the efficiency and accuracy of the tool.
Partially implemented future extensions include function models in Chebyshev bases, a calculus for differentiable functions, and Weierstrass approximation of general continuous functions.
Work is also in progress on more advanced classes of systems, notably on the evolution of nondeterministic hybrid systems described by differential inclusions~\citep{ZivanovicCollins2010}, and on model-checking linear temporal logic.
Theoretical work is in progress on the evolution of stiff continuous dynamics, the analysis of stochastic systems.

\subsection*{Acknowledgements}
 \includegraphics[height=1.8ex]{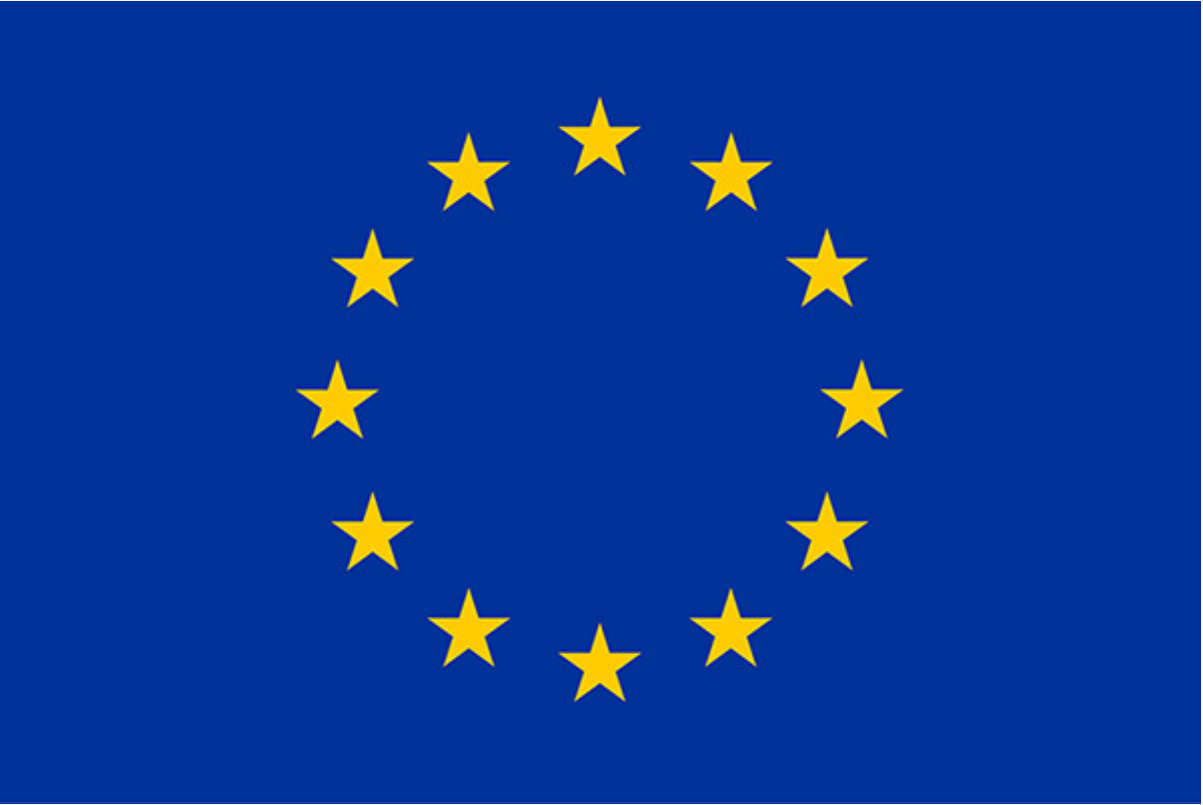}
This project has received funding from the European Union's Horizon 2020 research and innovation programme under the Marie Skłodowska-Curie grant agreement No 731143.

\small
\bibliographystyle{alphaurl}
\bibliography{ariadne-functioncalculus-lmcs}

\end{document}